\documentclass[a4paper,fleqn]{cas-dc}

\usepackage[numbers,sort&compress]{natbib}
\usepackage{amsmath,amssymb,bm}
\usepackage{booktabs}
\usepackage{graphicx}
\usepackage{microtype}
\usepackage{xcolor}

\newcommand{\ii}{\mathrm{i}}
\newcommand{\epszero}{\varepsilon_0}

\newcommand{\rev}[1]{#1}

\newcommand{\revgreen}[1]{#1}
\newcommand{\revblue}[1]{#1}

\begin{document}

\shorttitle{Dual-decomposition multi-GPU PIC for cylindrical plasmas}
\shortauthors{Y. Zhao, X. Chen, and Y. Chen}

\title[mode=title]{Dual-decomposition multi-GPU particle-in-cell method for cylindrical plasmas: Batched Fourier-mode multigrid and balanced particle slabs}

\author[1]{Yinjian Zhao}
\cormark[1]
\ead{zhaoyinjian@hit.edu.cn}

\author[1]{Xi Chen}

\author[1]{Yingjie Chen}

\affiliation[1]{organization={School of Energy Science and Engineering, Harbin Institute of Technology},
  city={Harbin},
  country={China}}

\cortext[1]{Corresponding author}

\begin{abstract}
\revblue{Three-dimensional electrostatic particle-in-cell (PIC) simulations
combine irregular particle operations with a globally coupled Poisson solve,
whose preferred parallel decompositions conflict.  We introduce a
dual-decomposition multi-GPU method: complete azimuthal Fourier modes are owned
during the field solve, whereas spatial slabs own particles.  Exact
full-spectrum diagonalization and batched matrix-free geometric multigrid keep
collectives outside the V-cycle, after which every GPU reconstructs the full
field.  Peer migration, cell reordering, warp-aggregated deposition, and
capacity-constrained dynamic cuts restore particle locality and balance.}
CPU/GPU Poisson solutions agree to relative $L_2$
errors below $6.9\times10^{-16}$; all 65 physical modes on eight V100 GPUs are
solved in 9.350~ms with relative errors below $8.2\times10^{-15}$.  For an
identical $512\times128\times400$ problem containing $707{,}788{,}800$
particles per species, the complete PIC loop reaches 0.149108~s per step and
strong-scales from five to eight GPUs with 91.81\% efficiency.  A separate
cross-resolution production study shows approximately threefold aggregate
particle-update and Poisson-cell throughput relative to a single RTX~5090
calculation; this measures refined-problem capability rather than a hardware
speedup or formal convergence order.
\end{abstract}


\begingroup
\hfuzz=125pt 
\maketitle
\endgroup

\section{Introduction}
\label{sec:introduction}

{\color{black}
Particle-in-cell (PIC) methods resolve kinetic plasma dynamics by coupling Lagrangian macro-particles to fields on an Eulerian mesh~\cite{BirdsallLangdon2005,HockneyEastwood1988}. Their numerical flexibility has enabled high-fidelity plasma calculations, but it also creates a heterogeneous workload: field gather and particle push stream irregular particle data, charge deposition introduces write conflicts, and the electrostatic Poisson equation globally couples the mesh. Recent GPU implementations therefore devote substantial effort to particle locality, on-chip data reuse, accelerator-aware deposition, asynchronous communication, and dynamic load balancing as well as arithmetic throughput~\cite{Xiong2024,Lee2025,Williams2025}.
}

{\color{black}
GPU PIC implementations have progressed from single-device kernels to scalable production frameworks such as PIConGPU, PSC, Smilei, WarpX, and GPU-enabled OSIRIS~\cite{Burau2010,Germaschewski2016,Derouillat2018,Myers2021,Lee2025}. Existing work has shown the value of particle tiling or sorting, accelerator-aware deposition, patch-based decomposition, and dynamic load balancing~\cite{Rossi2013,Germaschewski2016,Miller2021,Tan2022,Myers2021}. \revblue{A Hall thruster is an electric-propulsion device that ionizes and electrostatically accelerates a propellant; crossed electric and magnetic fields magnetize the electrons and drive an azimuthal Hall current~\cite{Hara2019,TaccognaGarrigues2019}.} Recent Hall-thruster calculations in WarpX further show both the promise of GPU acceleration and the scaling penalty of communication inside a distributed multigrid field solve~\cite{MarksGorodetsky2025}. These techniques are especially important when sources and losses create rapidly evolving particle distributions. However, a particle-optimal spatial decomposition is not necessarily field optimal: conventional mesh partitioning introduces halo exchanges and synchronization into an elliptic solve, while duplicating the complete field solve on every GPU wastes both memory and computation.
}

Cylindrical plasma devices offer an alternative. Periodicity in the azimuthal coordinate diagonalizes the angular operator and converts the three-dimensional Poisson problem into independent two-dimensional Fourier modes. Related quasi-cylindrical PIC algorithms exploit a small number of retained modes to reduce cost~\cite{Lehe2016}. The present objective is different: every discrete mode is retained so that the transformation is algebraically equivalent to the original three-dimensional finite-difference operator. The resulting shifted operator family is well suited to geometric multigrid, whose linear-complexity potential and parallel implementations are well established~\cite{Briggs2000,FalgoutJonesYang2006,Gholami2016}, but efficiently batching many complex mode solves and redistributing their spectra across a tightly coupled GPU node remain nontrivial.

{\color{black}
\revblue{In this work, we use a three-dimensional cylindrical Hall thruster plasma model as a demanding application for developing and evaluating the dual-decomposition method.} Fully kinetic Hall-thruster calculations are computationally demanding because electron magnetization, ion acceleration, sources, absorbing boundaries, and azimuthal instabilities must coexist over long physical times~\cite{Hara2019,TaccognaGarrigues2019,ZhaoZhao2026Review}. Two-dimensional and reduced-order PIC studies have clarified anomalous transport and electron-drift instabilities and provided community benchmarks~\cite{Lafleur2016,BoeufGarrigues2018,Villafana2021,Petronio2023,Reza2023,MarksGorodetsky2025}. \revblue{Recent work from our group has additionally quantified initialization and collision effects, introduced a three-dimensional Cartesian channel--plume code, and examined realistic magnetic-field components~\cite{Xie2024,Xie2025,Chen2025,Zhong2026}. Together with prior three-dimensional studies~\cite{Taccogna2006,TaccognaMinelli2018,Villafana2023}, these results demonstrate the importance of retaining coupled axial, radial, and azimuthal dynamics.} This application motivates an implementation that preserves the complete azimuthal spectrum and can sustain hundreds of millions of particles without sacrificing restartability.
}

{\color{black}
The unresolved design problem is therefore not merely to accelerate the field
or particle kernel in isolation, but to let the two subsystems use different
optimal ownership layouts without truncating the physical spectrum, repeatedly
redistributing the field, or duplicating the complete Poisson solve.  The
framework developed here makes that ownership transition an explicit part of
the time step while preserving the original discrete equations and particle
set.

This paper makes four contributions.  First, it derives an exact discrete,
full-spectrum azimuthal decomposition for a nonuniform cylindrical composite
domain; all physical modes share one protected-face, matrix-free multigrid
hierarchy.  Second, it assigns complete modes to GPUs and confines collective
communication to the two spectral redistribution boundaries, leaving every
multigrid V-cycle communication-free.  Third, it combines a complete
reconstructed field on each device with spatial particle ownership, direct
peer migration, cell reordering, and warp-aggregated cylindrical CIC, thereby
recovering particle locality without field halos.  Fourth, it introduces a
species- and capacity-constrained dynamic-cut policy that balances the physical
electron--ion work model while protecting both particle populations and their
communication buffers.  Together, these elements constitute an integrated
algorithm--system design rather than an isolated CUDA-kernel optimization; the
end-to-end experiments demonstrate numerical equivalence, sustained
hundreds-of-millions-particle operation, and 91.81\% strong-scaling efficiency
from five to eight GPUs.  Section~\ref{sec:physical_model} defines the physical
and numerical model; Sections~\ref{sec:mode_gmg} and
\ref{sec:particle_spatial} present the two decompositions; and
Section~\ref{sec:evaluation} evaluates accuracy, scaling, and production
behavior.
}

\section{Physical and numerical model}
\label{sec:physical_model}

\subsection{Electrostatic 3D3V particle model}

Electrons and singly charged xenon ions are represented by macro-particles in
three configuration-space and three velocity-space dimensions.  For a
particle of species $s\in\{e,i\}$, charge $q_s$, and mass $m_s$, the equations
of motion are
\begin{equation}
 \frac{\mathrm d\bm x_p}{\mathrm dt}=\bm v_p,
 \qquad
 \frac{\mathrm d\bm v_p}{\mathrm dt}
 =\frac{q_s}{m_s}\left(\bm E(\bm x_p,t)
 +\bm v_p\times\bm B(\bm x_p)\right).
 \label{eq:particle_motion}
\end{equation}
The electrostatic field is $\bm E=-\nabla\phi$, where
\begin{equation}
 -\nabla^2\phi=\frac{\rho}{\epszero},
 \qquad
 \rho=e(n_i-n_e).
 \label{eq:electrostatic_system}
\end{equation}
The prescribed magnetic field does not evolve.  Equation~\eqref{eq:particle_motion}
is advanced in leapfrog form with the standard Boris rotation~\cite{Boris1970,BirdsallLangdon2005}.
With $h_s=q_s\Delta t_s/(2m_s)$,
\begin{equation}
 \begin{split}
 \bm v^-&=\bm v^{n-1/2}+h_s\bm E^n,\qquad
 \bm t=h_s\bm B,\qquad
 \bm s=\frac{2\bm t}{1+|\bm t|^2},\\
 \bm v'&=\bm v^-+\bm v^-\times\bm t,\qquad
 \bm v^+=\bm v^-+\bm v'\times\bm s,\\
 \bm v^{n+1/2}&=\bm v^++h_s\bm E^n.
 \end{split}
 \label{eq:boris_update}
\end{equation}
The cylindrical drift is evaluated without a small-angle approximation.  If
$R=r+v_r\Delta t_s$ and $A=v_\alpha\Delta t_s$, then
\begin{equation}
 \begin{aligned}
 r^{n+1}&=\sqrt{R^2+A^2},\\
 \alpha^{n+1}&=\alpha^n+\operatorname{atan2}(A,R),\\
 z^{n+1}&=z^n+v_z\Delta t_s.
 \end{aligned}
 \label{eq:cylindrical_drift}
\end{equation}
Electrons are advanced every step with $\Delta t_e=\Delta t$.  Ions are
advanced once every $K$ electron steps with $\Delta t_i=K\Delta t$; their
previous deposited density is retained between ion updates.

{\color{black}
Electric fields are stored on cell faces and gathered with first-order
cloud-in-cell (CIC) weights.  Surviving particles deposit number density to
the eight surrounding nodes with the same weights, followed by a
volume-consistent projection to cell-centered charge.  This face-centered,
cylindrical-volume-consistent arrangement follows recent verification of
charge/current deposition and residual self-field behavior on nonuniform
three-dimensional cylindrical meshes~\cite{LiuZhouZhao2026}.  The particle
weight $W_s=n_0V_{\mathrm{active}}/N_s$ is common to all particles of one
species.  The complete step is: field gather and particle push;
absorbing-boundary and source processing; electron deposition and, on
subcycle steps, ion deposition; charge projection; Poisson solve; and
face-field reconstruction.  The multi-GPU ownership migration described in
Section~\ref{sec:particle_spatial} occurs after deposition and therefore
changes storage ownership, not the deposited physical state.
}

\subsection{Cylindrical composite domain and boundary conditions}

\begin{figure*}[t]
\centering
\includegraphics[width=0.92\textwidth]{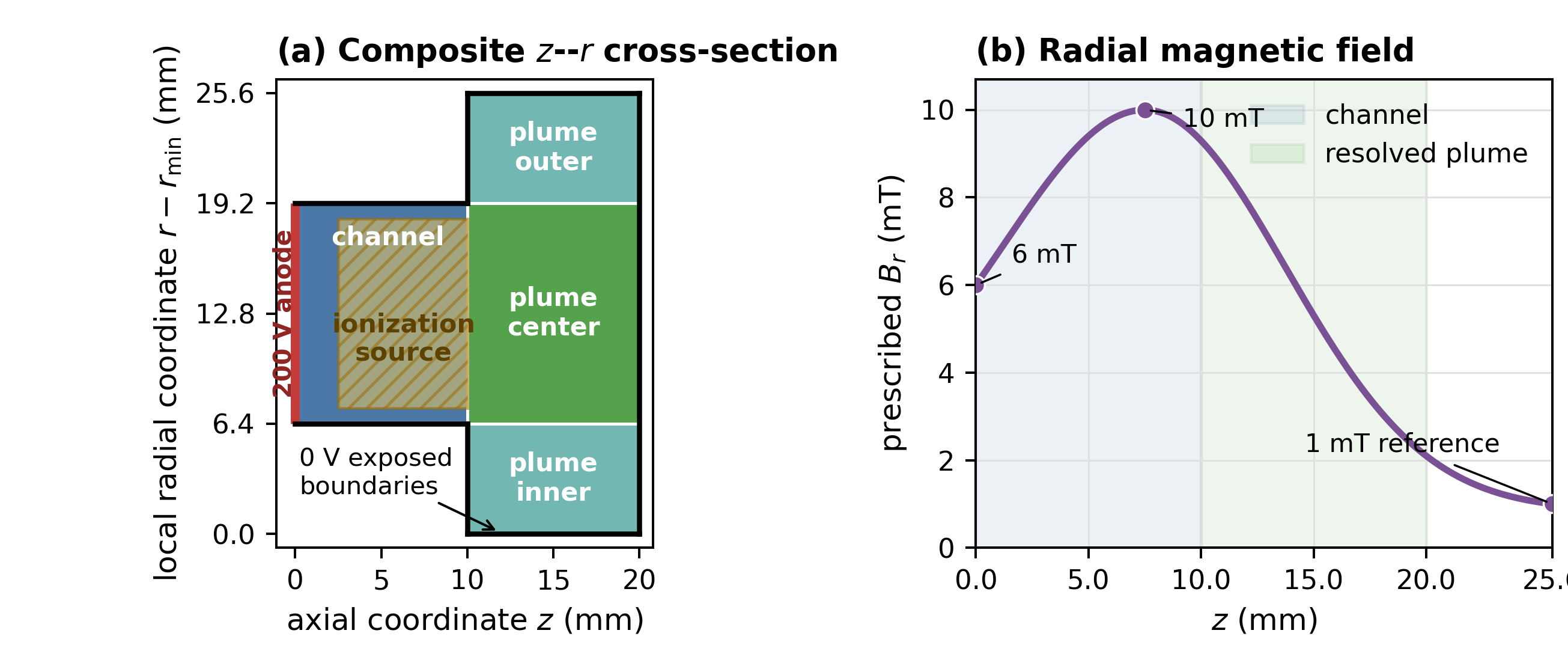}
\caption{\revgreen{Cylindrical Hall thruster plasma production case derived from
the archived input configuration.  (a) Active $z$--$r$ cross-section, with
axial distance shown horizontally and the local radial coordinate
$r-r_{\min}$ vertically.  The panel identifies the four nonoverlapping
computational patches, Dirichlet boundaries, and volumetric ionization
region.  The azimuthal direction is
periodic.  (b) Prescribed radial magnetic-field profile.  The coarse and fine
discretizations use the same physical geometry and differ only in mesh
spacing, time step, and particle sampling.}}
\label{fig:production_geometry}
\end{figure*}

The configuration space uses $(r,\alpha,z)$ coordinates with uniform,
periodic spacing in $\alpha$ and cell-centered finite volumes in the
$r$--$z$ plane.  The active cross-section is a union of nonoverlapping
rectangular patches.  Particle trajectories pass transparently across a
shared patch interface, whereas particles intersecting an exposed physical
face are absorbed and their number, charge, and kinetic energy are recorded.
The potential is periodic in $\alpha$ and satisfies the patch-face Dirichlet
conditions shown in Fig.~\ref{fig:production_geometry}.  This composite
description retains the channel--plume topology without filling inactive
corner regions.

\subsection{Production-case fields, initialization, and sources}
\label{sec:production_model}

{\color{black}
The evaluation uses the independently defined cylindrical Hall thruster
plasma case in Fig.~\ref{fig:production_geometry}.  The annular radius begins
at $r_{\min}=100$~m while the resolved radial width is only 25.6~mm.  This
large-radius convention follows a recent cylindrical PIC discretization
study~\cite{LiuZhouZhao2026} and facilitates comparison with an existing
three-dimensional Cartesian Hall-thruster calculation from the same research
group~\cite{Chen2025}.  Because the ratio of resolved radial width to
$r_{\min}$ is only $2.56\times10^{-4}$, curvature effects are deliberately
made negligible in the present production comparison; investigating finite
curvature is outside the present scope, although the solver itself retains
the cylindrical metric terms.  The azimuthal span is
$6.39918\times10^{-5}$~rad, corresponding to an arc length of approximately
6.4~mm at $r_{\min}$, and the axial length is 20~mm.
\revgreen{Writing $x=r-r_{\min}$ for the local radial coordinate, the channel
occupies $6.4\le x\le19.2$~mm and $0\le z\le10$~mm, while the plume occupies
$0\le x\le25.6$~mm and $10\le z\le20$~mm.}  The channel anode at $z=0$
is held at 200~V, and all other exposed field boundaries are held at 0~V.
}

The initial plasma is quasineutral with $n_e=n_i=5.0\times10^{16}$~m$^{-3}$.
Both species have zero mean drift and isotropic Maxwellian velocities with
$T_e=10$~eV and $T_i=0.5$~eV.  The species constants are
$q_e=-e$, $q_i=e$, $m_e=9.1093837015\times10^{-31}$~kg, and
$m_i=2.1801714\times10^{-25}$~kg.  The radial magnetic field is
\begin{equation}
 B_r(z)=a_\sigma
 \exp\!\left[-\frac{(z-z_p)^2}{2\sigma_B^2}\right]+b_\sigma,
 \qquad b_\sigma=B_p-a_\sigma,
 \label{eq:magnetic_profile}
\end{equation}
where $z_p=7.5$~mm and $\sigma_B=6.25$~mm.  Separate amplitudes on the two
sides of $z_p$ enforce $B_r(0)=6$~mT, $B_r(z_p)=10$~mT, and
$B_r(25.6\,\mathrm{mm})=1$~mT:
\begin{equation}
 \begin{aligned}
 a_-&=\frac{B_p-B_0}{1-\exp[-z_p^2/(2\sigma_B^2)]},\\
 a_+&=\frac{B_p-B_{\mathrm{end}}}
 {1-\exp[-(z_{\mathrm{end}}-z_p)^2/(2\sigma_B^2)]}.
 \end{aligned}
 \label{eq:magnetic_amplitudes}
\end{equation}

Ionization is prescribed as a time-independent volumetric pair-production
rate, uniform in $\alpha$,
\begin{equation}
 \mathcal S(x,z)=S_0 C(x;x_1,x_2)C(z;z_1,z_2),
 \label{eq:ionization_source}
\end{equation}
where the compact cosine envelope is
\begin{equation}
 C(\xi;a,b)=
 \begin{cases}
 \cos\!\left[\dfrac{\pi(\xi-(a+b)/2)}{b-a}\right],
       &a\le \xi\le b,\\[4pt]
 0, &\text{otherwise},
 \end{cases}
 \label{eq:cosine_envelope}
\end{equation}
The amplitude is $S_0=9.55\times10^{23}$~m$^{-3}$s$^{-1}$.
The source extends from $x_1=7.296$ to $x_2=18.304$~mm and from
$z_1=2.5$ to $z_2=10.0$~mm.
The expected macro-pair count is
$\Delta t\int\mathcal S\,r\,\mathrm dr\,\mathrm d\alpha\,\mathrm dz/W$;
its fractional part is sampled deterministically from the global step and
source seed.  New electrons and ions are colocated, with independent
Maxwellian velocities at 10 and 0.5~eV, respectively.

Cathode feedback is applied in the final axial cell layer.  If $N_e^c$ and
$N_i^c$ are its active electron and ion counts after the particle push, the
number of injected electrons is
\begin{equation}
 N_{\mathrm{inj}}^c=\max(N_i^c-N_e^c,0).
 \label{eq:cathode_feedback}
\end{equation}
Injected positions are uniform in cylindrical volume over that layer and
azimuthal interval.  Their velocities are sampled at 10~eV, with the axial
component directed into the domain.  Table~\ref{tab:production_physical_parameters}
collects the controls that are identical in the coarse and fine production
discretizations.  Initial-particle, ionization, and cathode random streams use
seeds 12345, 24680, and 97531, respectively.

\begin{table}[t]
\centering
\caption{Physical controls shared by both production discretizations.}
\label{tab:production_physical_parameters}
\small
\begin{tabular}{ll}
\toprule
Quantity & Value \\
\midrule
Domain $(x,r_{\min}\alpha,z)$ & $25.6\times6.4\times20$~mm$^3$ \\
Channel / plume axial extent & $10$ / $10$~mm \\
Anode / exposed boundaries & $200$ / $0$~V \\
Initial number density & $5.0\times10^{16}$~m$^{-3}$ \\
Electron / ion temperature & $10$ / $0.5$~eV \\
Magnetic field $(B_0,B_p,B_{\mathrm{end}})$ & $(6,10,1)$~mT \\
Ionization amplitude $S_0$ & $9.55\times10^{23}$~m$^{-3}$s$^{-1}$ \\
Ion-advance interval & $50$~ps \\
Poisson relative tolerance & $10^{-6}$ \\
\bottomrule
\end{tabular}
\end{table}

\section{Exact Fourier-mode decomposition and batched matrix-free multigrid}
\label{sec:mode_gmg}

Uniform periodic spacing in the azimuthal direction makes the angular part of the discrete Poisson operator circulant. It can therefore be diagonalized exactly by a discrete Fourier transform, while the nonuniform radial and axial meshes and the composite physical patches remain unchanged. The resulting two-dimensional equations differ only through a scalar angular eigenvalue. This shared structure is used to process all retained modes with one geometric multigrid hierarchy.

\begin{figure*}[t]
\centering
\IfFileExists{figs/fig01_fourier_mode_batched_gmg2.png}{%
  \includegraphics[width=\textwidth]{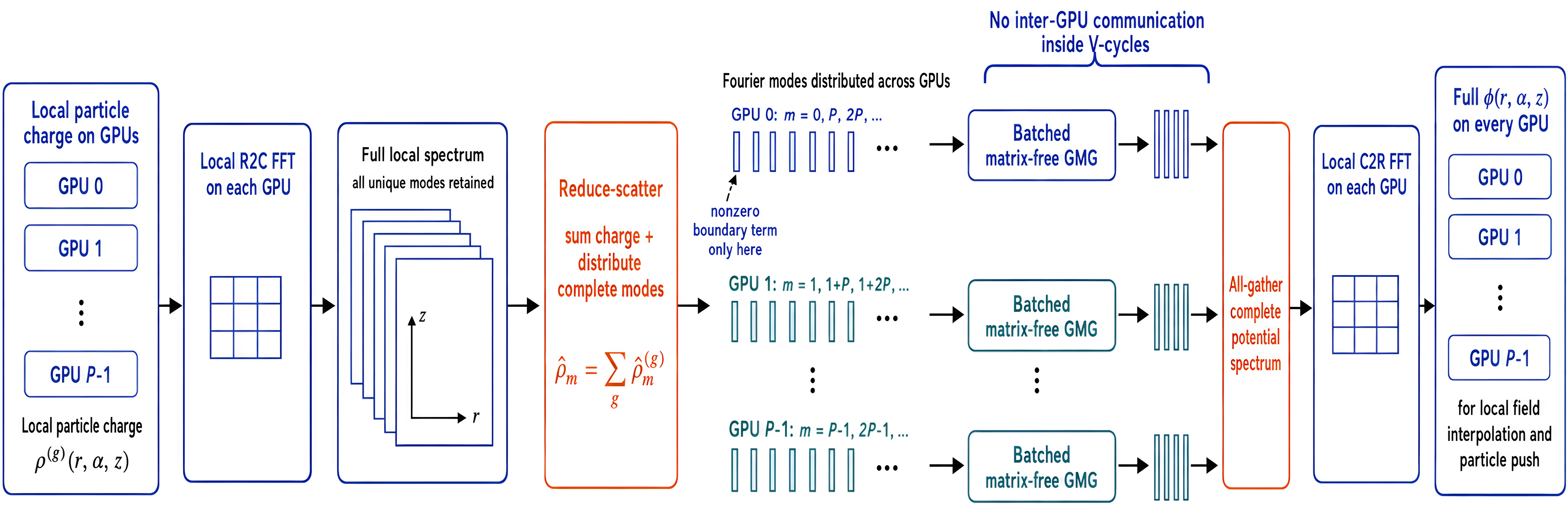}%
}{%
  \fbox{\parbox[c][0.205\textheight][c]{0.94\textwidth}{\centering
  Exact azimuthal R2C decomposition, mode redistribution, shared batched
  matrix-free GMG, spectrum collection, and C2R reconstruction.}}
}
\caption{Exact Fourier-mode decomposition and multi-GPU field-solver dataflow. Each GPU transforms its local particle charge, and reduce-scatter both sums the charge contributions and assigns complete modes to GPUs. Locally owned modes share one protected-face matrix-free GMG hierarchy and require no inter-GPU communication inside the V-cycles. All-gather and C2R then reconstruct the complete potential on every GPU.}
\label{fig:fourier_batched_gmg}
\end{figure*}

\subsection{Exact discrete azimuthal decoupling}

The electrostatic potential satisfies the cylindrical Poisson equation
\begin{equation}
-\left[
\frac{1}{r}\frac{\partial}{\partial r}
\left(r\frac{\partial\phi}{\partial r}\right)
+\frac{1}{r^2}\frac{\partial^2\phi}{\partial\alpha^2}
+\frac{\partial^2\phi}{\partial z^2}
\right]
=\frac{\rho}{\epszero}.
\label{eq:cylindrical_poisson}
\end{equation}
Here, $r$ is the radial coordinate, $\alpha$ is the periodic azimuthal coordinate, $z$ is the axial coordinate, $\phi(r,\alpha,z)$ is the electric potential, $\rho(r,\alpha,z)$ is the charge density, and $\epszero$ is the vacuum permittivity.

Let the periodic azimuthal interval have length $L_\alpha$ and contain $N_\alpha$ uniformly spaced cells, with $\Delta\alpha=L_\alpha/N_\alpha$ and $\alpha_j=\alpha_0+j\Delta\alpha$. The indices $i$, $j$, and $k$ identify radial, azimuthal, and axial cells, respectively; $m\in\{0,\ldots,N_\alpha-1\}$ is the Fourier-mode index; a hat denotes a Fourier coefficient; and $\ii=\sqrt{-1}$ is the imaginary unit. For any grid function $f$, its discrete Fourier coefficient is
\begin{equation}
\widehat{f}_m(r_i,z_k)
=\frac{1}{N_\alpha}
\sum_{j=0}^{N_\alpha-1}
f(r_i,\alpha_j,z_k)
\exp\left(-\frac{2\pi\ii m j}{N_\alpha}\right).
\label{eq:azimuthal_dft}
\end{equation}

The angular second derivative in the original seven-point finite-difference operator is
\begin{equation*}
(\delta_{\alpha\alpha}f)_j
=\frac{f_{j+1}-2f_j+f_{j-1}}{\Delta\alpha^2},
\end{equation*}
where periodic indexing identifies $f_{-1}$ with $f_{N_\alpha-1}$ and $f_{N_\alpha}$ with $f_0$. Define the discrete Fourier vector $v_j^{(m)}=\exp(\ii\theta_m j)$, where $\theta_m=2\pi m/N_\alpha$. Applying the negative discrete angular operator to this vector gives
\begin{align*}
-\delta_{\alpha\alpha}v_j^{(m)}
&=\frac{2-e^{\ii\theta_m}-e^{-\ii\theta_m}}
        {\Delta\alpha^2}v_j^{(m)}\\
&=\frac{4}{\Delta\alpha^2}
  \sin^2\left(\frac{\theta_m}{2}\right)v_j^{(m)}.
\end{align*}
Thus the eigenvalue of the matrix that is actually used to discretize $-\partial^2/\partial\alpha^2$ is
\begin{equation}
\lambda_m
=\frac{4}{\Delta\alpha^2}
\sin^2\left(\frac{\pi m}{N_\alpha}\right).
\label{eq:discrete_azimuthal_eigenvalue}
\end{equation}
The phrase \emph{exact discrete eigenvalue} means that Eq.~\eqref{eq:discrete_azimuthal_eigenvalue} diagonalizes the finite-difference matrix without approximation. It is distinct from the continuous Fourier eigenvalue $k_m^2$, where $k_m=2\pi m/L_\alpha$ is the continuous angular wave number. For fixed $m$, $\lambda_m$ approaches $k_m^2$ as $\Delta\alpha$ tends to zero, but using $k_m^2$ at finite resolution would no longer reproduce the original three-dimensional discrete operator exactly.

To show how Eq.~\eqref{eq:discrete_azimuthal_eigenvalue} enters Eq.~\eqref{eq:cylindrical_poisson}, write the inverse transform of the potential as
\begin{equation*}
\phi(r_i,\alpha_j,z_k)
=\sum_{m=0}^{N_\alpha-1}
\widehat{\phi}_m(r_i,z_k)e^{\ii\theta_mj},
\end{equation*}
and use the analogous expansion for $\rho$. Radial and axial derivatives do not act on $e^{\ii\theta_mj}$, whereas the discrete angular term satisfies $-\delta_{\alpha\alpha}e^{\ii\theta_mj}=\lambda_m e^{\ii\theta_mj}$. Equating the coefficient of each Fourier vector therefore gives
\begin{equation*}
-\frac{1}{r}\frac{\partial}{\partial r}
 \left(r\frac{\partial\widehat{\phi}_m}{\partial r}\right)
-\frac{\partial^2\widehat{\phi}_m}{\partial z^2}
+\frac{\lambda_m}{r^2}\widehat{\phi}_m
=\frac{\widehat{\rho}_m}{\epszero}.
\end{equation*}
Multiplying this equation by $r$ yields the two-dimensional equation used by the mode solver,
\begin{equation}
-\frac{\partial}{\partial r}
\left(r\frac{\partial\widehat{\phi}_m}{\partial r}\right)
-r\frac{\partial^2\widehat{\phi}_m}{\partial z^2}
+\frac{\lambda_m}{r}\widehat{\phi}_m
=\frac{r}{\epszero}\widehat{\rho}_m.
\label{eq:mode_poisson}
\end{equation}
Here, $\widehat{\phi}_m(r,z)$ and $\widehat{\rho}_m(r,z)$ are the potential and charge-density coefficients of mode $m$. For real-valued physical fields and even $N_\alpha$, conjugate symmetry leaves $N_m=N_\alpha/2+1$ unique real-to-complex (R2C) modes. All $N_m$ modes are retained, so the transformation changes the solver representation but does not truncate the physics.

\subsection{A shared shifted-operator family}

Equation~\eqref{eq:mode_poisson} is discretized on the packed active cells of the composite $r$--$z$ cross-section. Let $q=(i,k)$ denote one active cell, let $r_i$ and $z_k$ be its center coordinates, and let $\Delta r_i$ and $\Delta z_k$ be its radial and axial widths. The angular term contributes the cell coefficient
\begin{equation}
\mu_q=\frac{\Delta r_i\Delta z_k}{r_i}.
\label{eq:mode_mass_coefficient}
\end{equation}
The symbol $\mu_q$ is therefore a geometry factor, not a fitted physical parameter.

Let $u_{q,m}=\widehat{\phi}_m(r_i,z_k)$ be the unknown potential of mode $m$ in cell $q$, and collect all cell values into the vector $u_m$. Let $A_m$ denote the discrete operator for mode $m$, $\mathcal{N}(q)$ the set of active radial and axial neighbors of cell $q$, and $q'$ one member of this set. The positive coefficient $G_{qq'}$ is the finite-volume face conductance between $q$ and $q'$, while $D_q^{(0)}$ is the sum of the mode-independent interior and Dirichlet-face conductances attached to $q$. Integrating Eq.~\eqref{eq:mode_poisson} over cell $q$ produces the matrix-free stencil
\begin{equation}
(A_m u_m)_q
=\left(D_q^{(0)}+\lambda_m\mu_q\right)u_{q,m}
-\sum_{q'\in\mathcal{N}(q)}G_{qq'}u_{q',m}.
\label{eq:matrix_free_mode_operator}
\end{equation}
In direct terms, Eq.~\eqref{eq:matrix_free_mode_operator} multiplies the potential in the current cell by its diagonal coefficient and subtracts one conductance-weighted potential for each active neighbor. No sparse matrix is stored on the GPU.

Only the diagonal term $\lambda_m\mu_q$ changes with the Fourier mode. If $A_0$ denotes Eq.~\eqref{eq:matrix_free_mode_operator} with $\lambda_0=0$, and if $M$ is the diagonal matrix whose $q$th entry is $\mu_q$, then
{\color{black}
\begin{equation}
A_m=A_0+\lambda_m M,
\qquad
M=\operatorname{diag}(\mu_q).
\label{eq:shifted_operator_family}
\end{equation}
}
Equation~\eqref{eq:shifted_operator_family} is a compact statement of reuse: every mode has the same cells, neighbor lists, off-diagonal conductances, physical interfaces, and multigrid transfer maps. The value $\lambda_m$ merely adds a different positive shift to each cell diagonal. Appendix~\ref{app:fv_shifted_operator} gives the individual radial, axial, and boundary coefficients that form $D_q^{(0)}$ and $G_{qq'}$.

Let $b_{q,m}$ be the algebraic right-hand-side entry, let $\widehat{\rho}_{q,m}=\widehat{\rho}_m(r_i,z_k)$ be the transformed charge density, let $s_q$ be its finite-volume scale, and let $b_q^D$ be the contribution from nonzero Dirichlet faces. With $\delta_{m0}$ denoting the Kronecker delta, the discrete right-hand side in cell $q$ is
\begin{equation}
b_{q,m}=s_q\widehat{\rho}_{q,m}
+b^{D}_{q}\,\delta_{m0},
\qquad
s_q=\frac{r_i\Delta r_i\Delta z_k}{\epszero}.
\label{eq:mode_rhs}
\end{equation}
The factor $\delta_{m0}$ follows from the assumed azimuthal invariance of the boundary values: a constant boundary has only a zero Fourier mode. Azimuthally varying boundary data would instead contribute its corresponding Fourier coefficients to the affected modes.

Patch boundaries and interfaces are retained as protected faces on every coarse level, so coarsening never merges cells across a physical discontinuity or a geometric gap. The hierarchy therefore reuses geometry without erasing the composite-domain topology.

\subsection{Complex batched GPU execution}

Suppose one GPU owns $B$ Fourier modes $m_0,\ldots,m_{B-1}$. Let the lane index $\ell\in\{0,\ldots,B-1\}$ select one locally owned mode and let $B_{\mathrm{pad}}\geq B$ be the padded mode extent used to align each cell row. The operators $\operatorname{Re}$ and $\operatorname{Im}$ select real and imaginary components, respectively. The complex unknowns are stored in a cell-major, mode-fast array,
\begin{equation}
\texttt{value}\,[qB_{\mathrm{pad}}+\ell]
=\left(
\operatorname{Re}u_{q,m_\ell},
\operatorname{Im}u_{q,m_\ell}
\right).
\label{eq:mode_fast_layout}
\end{equation}
One double-precision complex pair is stored per entry. Neighboring GPU lanes process different modes of the same $r$--$z$ cell, so their accesses are contiguous and the cell geometry can be reused across the batch. The same layout is used by smoothing, residual evaluation, restriction, prolongation, and the coarse-grid solve.

This execution differs from constructing one general sparse-matrix hierarchy per mode and solving the real and imaginary systems sequentially. The batched solver uploads one protected-face hierarchy, keeps all level vectors resident on the device, and applies the mode-dependent shift from Eq.~\eqref{eq:shifted_operator_family} inside the stencil kernel. \rev{The independent reference and fallback uses HYPRE's structured-grid parallel semicoarsening multigrid solver, specifically the \texttt{HYPRE\_StructPFMG} setup and solve interface, once for each real or imaginary mode component~\cite{FalgoutJonesYang2006}; it is not the production batched data layout.}

\subsection{Multi-GPU communication boundary}

With $P$ GPUs, let $g\in\{0,\ldots,P-1\}$ be a GPU index. GPU $g$ initially deposits the charge generated by its local particles and computes a full local R2C spectrum $\widehat{\rho}^{(g)}_m$, where the superscript $(g)$ identifies the contributing GPU. A reduce-scatter operation forms the global coefficient
\begin{equation}
\widehat{\rho}_m
=\sum_{g=0}^{P-1}\widehat{\rho}^{(g)}_m
\label{eq:global_charge_spectrum}
\end{equation}
and distributes complete modes according to a cyclic ownership rule. Denoting the set of complete modes assigned to GPU $g$ by $\mathcal{M}_g$, this rule is
\begin{equation}
\mathcal{M}_g
=\left\{m\;\middle|\;
0\leq m<N_m,\;m\bmod P=g
\right\}.
\label{eq:cyclic_mode_ownership}
\end{equation}

Each GPU advances its local batch through the GMG V-cycles. No halo exchange or collective communication occurs inside a V-cycle because a complete two-dimensional mode is owned by exactly one GPU. Only the owner of $m=0$ applies the nonzero boundary term in Eq.~\eqref{eq:mode_rhs}. After convergence or a prescribed number of cycles, an all-gather reconstructs the complete potential spectrum on every GPU, and a local complex-to-real transform produces the full $\phi(r,\alpha,z)$ required by that GPU's particles.

Figure~\ref{fig:fourier_batched_gmg} summarizes the dataflow and shared hierarchy. The design is communication avoiding rather than communication free: synchronization is concentrated in the two spectral collectives surrounding the solve, while the V-cycle itself remains device local.

\section{Spatial particle decomposition with locality recovery and capacity-aware balancing}
\label{sec:particle_spatial}

The particle and field subsystems use different ownership rules. During the field solve, each GPU owns a subset of the azimuthal Fourier modes. After the spectral all-gather and inverse transform, however, every GPU holds the complete three-dimensional electric field. Particles can therefore be assigned by physical position without introducing remote field interpolation. In the present implementation, each particle belongs to one contiguous axial slab; communication is required only when a particle crosses a slab boundary or when the slab cuts are deliberately changed.

\begin{figure*}[t]
\centering
\IfFileExists{figs/fig02_particle_spatial_pipeline.png}{%
  \includegraphics[width=\textwidth]{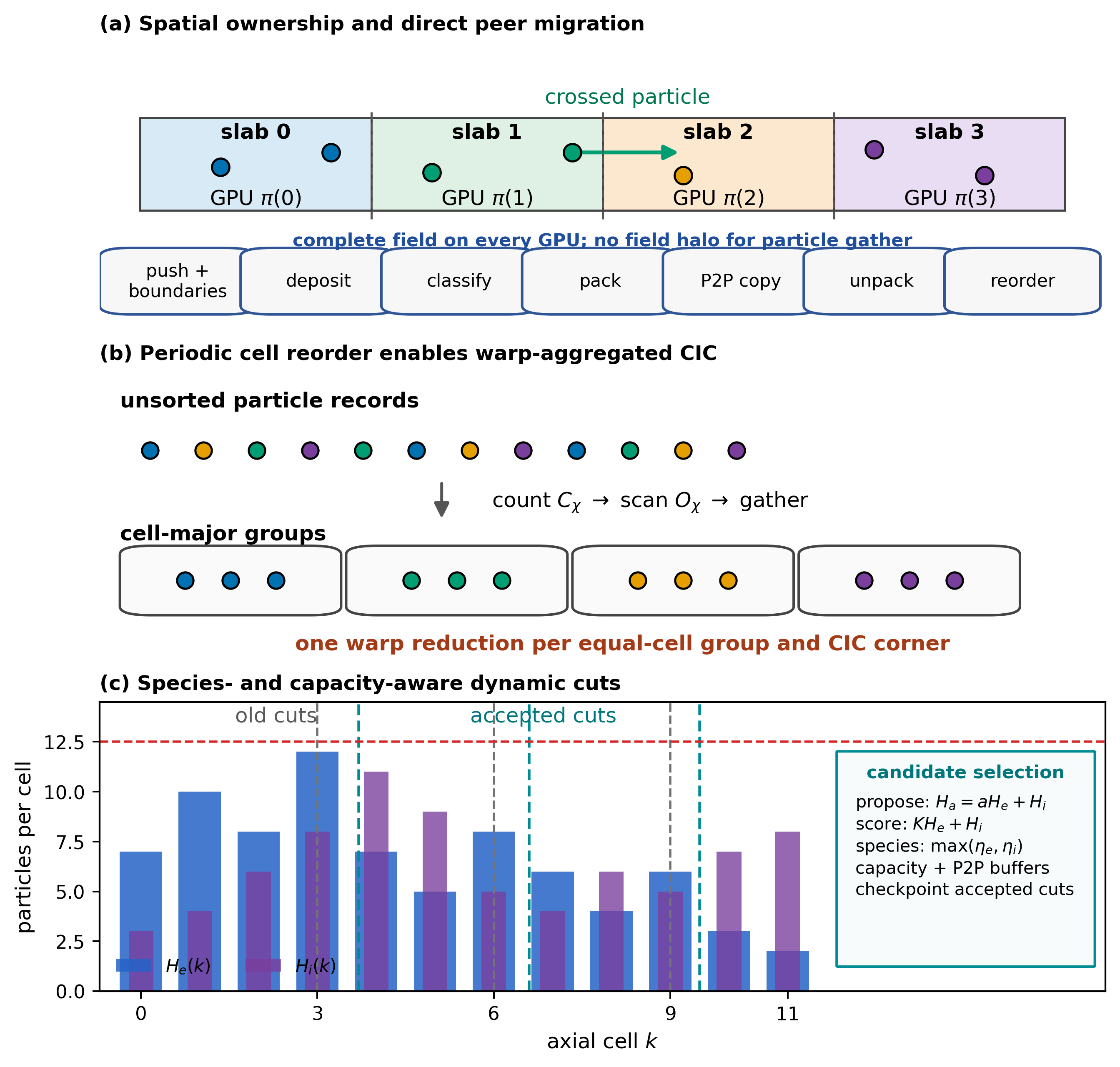}%
}{%
  \fbox{\parbox[c][0.215\textheight][c]{0.94\textwidth}{\centering
  Particle-step order, axial ownership, direct peer migration, cell
  reordering, warp-aggregated CIC, and dynamic capacity-aware cuts.}}
}
\caption{\revgreen{Particle-processing path.  (a) Axial
slabs define spatial ownership; particles deposit at their updated physical
position before crossed particles are packed and transferred directly to the
new owner.  (b) Periodic counting sort restores a dense cell-major layout and
creates equal-cell warp groups for aggregated CIC.  (c) Global electron and
ion histograms propose new integer cuts; a proposal is accepted only when it
improves balance and satisfies species-capacity and temporary-memory checks.}}
\label{fig:particle_spatial_pipeline}
\end{figure*}

\subsection{Particle-step order and ownership invariant}

The order of operations is important. At electron time level \(n\), the complete electric field is available on every GPU. Each device then (i) gathers the field and pushes its particles, (ii) applies particle sources and physical boundary conditions, (iii) deposits charge on its local copy of the full mesh, (iv) migrates particles whose new positions belong to another axial slab, and (v) performs a cell reorder when the prescribed reorder interval is reached. The deposited arrays are subsequently combined in the mode-decomposed Poisson solve, and the reconstructed field is used at time level \(n+1\).

Depositing before migration does not double count or omit a particle. Let \(\mathcal P_g^{-}\) and \(\mathcal P_g^{+}\) be the active particle sets stored on GPU \(g\) immediately before and after migration, respectively, and let \(\mathcal D(\mathcal P)\) denote the additive CIC deposition of a particle set \(\mathcal P\). Migration changes the storage owner but not the global particle set, so
\begin{equation*}
\biguplus_{g=0}^{P-1}\mathcal P_g^{-}
=\biguplus_{g=0}^{P-1}\mathcal P_g^{+},
\qquad
\sum_{g=0}^{P-1}\mathcal D(\mathcal P_g^{-})
=\sum_{g=0}^{P-1}\mathcal D(\mathcal P_g^{+}),
\end{equation*}
where \(P\) is the number of GPUs and \(\biguplus\) denotes a disjoint union. Thus, the pre-migration deposition is the same global deposition that would be obtained after migration, apart from the permitted change in floating-point summation order. Migration prepares correct ownership for the next particle step. Appendix~\ref{app:particle_pipeline} illustrates this invariant with a boundary-crossing particle.

\subsection{Axial ownership and direct particle migration}

Let \(N_z\) be the number of axial cells and let \(c_g\) be the first cell owned by logical GPU \(g\in\{0,\ldots,P-1\}\). The ordered integer cut vector is
\begin{equation}
0=c_0<c_1<\cdots<c_P=N_z.
\label{eq:particle_z_cuts}
\end{equation}
For particle \(p\), let \(i_z(p)\) be the axial-cell index cached after the particle push. Its unique logical owner is
\begin{equation}
g(p)=g
\quad\Longleftrightarrow\quad
c_g\leq i_z(p)<c_{g+1}.
\label{eq:particle_z_owner}
\end{equation}
A permutation \(\pi\) maps logical slab \(g\) to physical device \(\pi(g)\). The geometrical ordering of slabs can consequently be preserved while the physical assignment follows the available peer-to-peer topology.

Let \(g_{\mathrm{old}}(p)\) be the owner before migration and let \(h\) be a possible destination. The particles sent from logical GPU \(g\) to logical GPU \(h\) form
\begin{equation}
\mathcal O_{g\rightarrow h}
=\left\{p\mid g_{\mathrm{old}}(p)=g,
                 \;g(p)=h,\;h\neq g\right\}.
\label{eq:outgoing_particle_set}
\end{equation}
Migration consists of four explicit stages. First, a device pass counts \(\lvert\mathcal O_{g\rightarrow h}\rvert\) for every destination \(h\), where vertical bars denote set cardinality. Second, prefix offsets are formed and every outgoing particle is packed into a destination-contiguous buffer. The packed item contains the complete phase-space coordinates, velocity components, cached cell indices, and composite-domain region identifier; the receiver therefore does not reconstruct partial particle state. Third, asynchronous CUDA peer copies move the contiguous destination segments directly between physical devices. Finally, the receiver fills recycled particle slots before extending its active high-water mark. Cell reordering is performed only after this transfer, because the reorder assumes that every retained active particle belongs to the local slab.

\subsection{Periodic cell reordering}

Axial ownership does not by itself provide cell locality. Particle pushing, absorption, source insertion, and migration progressively scramble the structure-of-arrays storage. A periodic counting sort restores a dense, cell-major order without changing the represented particle set.

Let \(i_r(p)\), \(i_\alpha(p)\), and \(i_z(p)\) be the radial, azimuthal, and axial cell indices of particle \(p\), respectively. For active radial--axial cell \((i_r,i_z)\), let \(B_{i_r,i_z}\) be the first valid packed key and let \(S_{i_r,i_z}\) be the topology-dependent stride between adjacent azimuthal cells. The complete packed cell key is
\begin{equation}
\kappa(p)
=B_{i_r(p),i_z(p)}
+i_\alpha(p)S_{i_r(p),i_z(p)}.
\label{eq:particle_cell_key}
\end{equation}
The base \(B_{i_r,i_z}\) and stride \(S_{i_r,i_z}\) come from the composite-domain topology; Eq.~\eqref{eq:particle_cell_key} therefore preserves inactive gaps and patch interfaces instead of treating the mesh as a rectangular box.

For each valid key \(\chi\), the counting pass obtains a population \(C_\chi\). An exclusive scan gives the first destination position \(O_\chi=\sum_{\psi<\chi}C_\psi\), where \(\psi\) is another valid key ordered before \(\chi\). A scatter assigns one position in the interval \([O_\chi,O_\chi+C_\chi)\) to every particle with key \(\chi\), and a final gather applies the resulting permutation to every structure-of-arrays field. Inactive and migrated entries are omitted, leaving a dense active prefix. The sort is invoked periodically rather than at every time step so that its cost is amortized while the reordered layout benefits several subsequent gathers and depositions. A six-particle example is given in Appendix~\ref{app:particle_pipeline}.

\subsection{Warp-aggregated cylindrical CIC deposition}

For particle \(p\), let \((\xi_p,\eta_p,\zeta_p)\in[0,1]^3\) be its normalized position within the radial, azimuthal, and axial extents of the cached cell. For coordinate direction \(\ell\in\{r,\alpha,z\}\), the lower- and upper-node cloud-in-cell (CIC) weights are denoted by \(w_\ell^{(0)}\) and \(w_\ell^{(1)}\), respectively:
\begin{equation}
\begin{aligned}
w_r^{(0)}&=1-\xi_p, & w_r^{(1)}&=\xi_p,\\
w_\alpha^{(0)}&=1-\eta_p, & w_\alpha^{(1)}&=\eta_p,\\
w_z^{(0)}&=1-\zeta_p, & w_z^{(1)}&=\zeta_p.
\end{aligned}
\label{eq:cic_weights}
\end{equation}
Let \(\bm d=(d_r,d_\alpha,d_z)\in\{0,1\}^3\) select one of the eight cell corners, let \(\nu(\bm d)\) be the global node at that corner, let \(W_p\) be the macro-particle weight, let \(V_{\nu(\bm d)}\) be the cylindrical control volume associated with the node, and let \(n_{\nu(\bm d)}\) be the deposited nodal number density. Conventional CIC performs
\begin{equation}
n_{\nu(\bm d)}\mathrel{+}=
\frac{W_p}{V_{\nu(\bm d)}}
w_r^{(d_r)}w_\alpha^{(d_\alpha)}w_z^{(d_z)}
\label{eq:particle_cic_update}
\end{equation}
for all eight choices of \(\bm d\), requiring eight double-precision atomic additions per particle.

The cell-major order makes particles with the same cell label contiguous. Let \(R(p)\) denote the composite-domain region containing particle \(p\), and define the exact cell label
\begin{equation}
\chi_p=\left(R(p),i_z(p),i_\alpha(p),i_r(p)\right).
\label{eq:warp_cell_label}
\end{equation}
Threads in each warp are partitioned by equal values of \(\chi_p\). For one warp-local group \(G_\chi\) and one corner \(\bm d\), the lane contributions are reduced in registers to
\begin{equation}
\Delta n_{\nu(\bm d),G_\chi}
=\frac{1}{V_{\nu(\bm d)}}
\sum_{p\in G_\chi}W_p
w_r^{(d_r)}w_\alpha^{(d_\alpha)}w_z^{(d_z)},
\label{eq:warp_aggregated_cic}
\end{equation}
where \(\Delta n_{\nu(\bm d),G_\chi}\) is the total contribution of the group to the selected node. Only the group leader performs the global atomic addition. A group containing \(q\) particles therefore replaces \(8q\) global atomics by at most eight, one for each corner. All particle weights and all eight CIC corners remain present; only the order of addition changes. Appendix~\ref{app:particle_pipeline} gives a concrete three-particle example.

\subsection{Species- and capacity-aware dynamic cuts}

Equal-length slabs can become imbalanced as ionization, cathode injection, and boundary loss reshape the particle distribution. Rebalancing is evaluated only at an interval that is compatible with both the cell-reorder period and the ion subcycle. At that point, both species have been migrated under the current cuts and their local arrays are dense.

Let \(s\in\{e,i\}\) identify electrons (\(e\)) or ions (\(i\)), and let \(\mathcal A_s(k)\) be the set of active particles of species \(s\) in axial cell \(k\). The global axial histogram is
\begin{equation}
\begin{aligned}
\mathcal A_s(k)
  &=\left\{p\in s\mid p\ \text{is active},\ i_z(p)=k\right\},\\
H_s(k)&=\#\mathcal A_s(k),
\qquad k=0,\ldots,N_z-1,
\end{aligned}
\label{eq:species_z_histogram}
\end{equation}
where \(H_s(k)\) is the active-particle count in axial cell \(k\), and \(\#\) denotes the number of elements in a set. If ions are advanced once every \(K\) electron steps, the particle work accumulated during one ion subcycle is modeled by
\begin{equation}
H_{\mathrm{work}}(k)=K H_e(k)+H_i(k),
\label{eq:particle_work_histogram}
\end{equation}
where \(K\) is the ion-subcycle factor.

For a trial cut vector \(\mathcal C=(c_1,\ldots,c_{P-1})\), let \(x\in\{e,i,\mathrm{work}\}\) select a histogram. The load assigned to GPU \(g\) and its max-to-average imbalance ratio are
\begin{equation}
L_g^{(x)}(\mathcal C)
=\sum_{k=c_g}^{c_{g+1}-1}H_x(k),
\qquad
\eta_x(\mathcal C)
=\frac{P\max_g L_g^{(x)}(\mathcal C)}
       {\sum_g L_g^{(x)}(\mathcal C)}.
\label{eq:particle_load_ratio}
\end{equation}
Here \(L_g^{(x)}\) is the load of quantity \(x\) in slab \(g\), and \(\eta_x=1\) represents perfect balance.

A cut obtained from \(H_{\mathrm{work}}\) alone can improve the predicted running time while placing too many particles of one species on a GPU. Candidate cuts are therefore generated from a small family of auxiliary histograms
\begin{equation}
H_a(k)=aH_e(k)+H_i(k),
\qquad a=1,\ldots,K,
\label{eq:particle_cut_candidates}
\end{equation}
where the integer \(a\) is only a proposal weight. Every proposed cut is evaluated with the physical work weight \(K\), not with \(a\). Its worst per-species imbalance is
\begin{equation}
\eta_{\mathrm{species}}(\mathcal C)
=\max\!\left(\eta_e(\mathcal C),\eta_i(\mathcal C)\right).
\label{eq:worst_species_ratio}
\end{equation}
During an ordinary periodic check, a candidate must reduce \(\eta_{\mathrm{work}}\) relative to the current cuts and keep \(\eta_{\mathrm{species}}\) below a prescribed limit. If a device approaches its particle-array capacity, the policy changes priority: the candidate must first produce a strict improvement in \(\eta_{\mathrm{species}}\), after which candidates are ranked by their true work imbalance. Appendix~\ref{app:particle_pipeline} uses an eight-cell histogram to show why the auxiliary values \(a<K\) can yield a safer cut than the work-only choice \(a=K\).

Let \(M_{g,s}\) be the allocated particle capacity for species \(s\) on GPU \(g\), and let \(f_{\mathrm{stop}}\in(0,1)\) be the reserved capacity fraction. An accepted cut \(\mathcal C_{\mathrm{new}}\) must satisfy
\begin{equation}
L_g^{(s)}(\mathcal C_{\mathrm{new}})
\leq(1-f_{\mathrm{stop}})M_{g,s}
\label{eq:particle_capacity_constraint}
\end{equation}
for every GPU and both species. The implementation also predicts the one-time send and receive volumes implied by the old and new owners. A candidate is rejected if the temporary peer buffers would violate the free-memory reserve of any device.

When a cut change is accepted, both species are passed through the existing migration path, their arrays are reordered and compacted, and the new cuts and physical-device permutation are written to the checkpoint metadata. A restarted simulation installs this layout before loading the device shards. Dynamic particle ownership is consequently part of the reproducible simulation state rather than an unrecorded performance decision.

The complete path is summarized in Fig.~\ref{fig:particle_spatial_pipeline}. Migration changes storage ownership, and reordering changes storage order. Warp aggregation changes the arithmetic reduction order. None of these operations changes the PIC particle set or the eight-corner CIC discretization.

\section{Experimental evaluation}
\label{sec:evaluation}

The evaluation addresses four questions.  First, does the matrix-free batched
solver reproduce a trusted full three-dimensional solution while retaining all
azimuthal modes?  Second, does mode ownership reduce the end-to-end Poisson
time on a tightly connected multi-GPU node?  Third, does the complete PIC loop
strong-scale without changing the physical result when particle ownership is
changed?  Finally, does the implementation sustain a long, refined
cylindrical \rev{Hall thruster plasma} calculation while preserving the macroscopic behavior
of its coarse-grid counterpart?  The experiments below separate same-problem
strong scaling from the cross-generation, cross-resolution production
comparison.

\subsection{Platforms, workloads, and metrics}
\label{sec:evaluation_protocol}

Table~\ref{tab:evaluation_matrix} summarizes the archived data sets.  The V100
node contained eight Tesla V100-SXM2 GPUs with 32~GiB per device.  Its fully
connected CUDA peer-access matrix included bonded NVLink pairs, and collectives
were executed with NCCL.  The RTX~5090 environment used driver 580.82.07,
CUDA 12.8, and double-precision HYPRE 3.1.0 for the independent structured-grid
reference.  The V100 environment used driver 535.216.01 and CUDA 12.2, with
NCCL version 2.31.2.
The CPU reference used 64 host cores; the processor model was not retained in
the archived bundle, so CPU results are reported only as a numerical and
within-run timing reference rather than a portable hardware benchmark.

\begin{table*}[t]
\centering
\caption{\rev{Evaluation matrix.  Particle counts are per species.  The
wrapped-column layout avoids scaling the table text below the manuscript
font.}}
\label{tab:evaluation_matrix}
\small
{\color{black}
\setlength{\tabcolsep}{2.2pt}
\begin{tabular}{@{}p{0.16\textwidth}p{0.14\textwidth}p{0.20\textwidth}
                p{0.12\textwidth}p{0.18\textwidth}p{0.15\textwidth}@{}}
\toprule
Experiment & Platform & Grid & Particles & Timing protocol & Purpose \\
\midrule
Poisson CPU/GPU reference
 & 64 CPU cores / 1 RTX 5090
 & $256\!\times\!128\!\times\!256$, $512\!\times\!256\!\times\!512$
 & -- & 2 warmups, $5$--$13$ samples & reference accuracy and single-GPU speed \\
Poisson mode matrix
 & 2--8 V100
 & $512\!\times\!128\!\times\!400$
 & -- & 3 rounds, 10 warmups + 30 samples & spectral redistribution and mode scaling \\
Full-PIC scaling
 & 5--8 V100
 & $512\!\times\!128\!\times\!400$
 & $707{,}788{,}800$
 & 3 rounds of 1200 steps; last 1000 timed & same-problem strong scaling \\
Cylindrical production, coarse
 & 1 RTX 5090
 & $256\!\times\!64\!\times\!200$
 & $88{,}473{,}600$
 & archived production history & baseline physics and throughput \\
Cylindrical production, fine
 & 8 V100
 & $512\!\times\!128\!\times\!400$
 & $314{,}572{,}800$
 & archived production history & refined physics and multi-GPU throughput \\
\bottomrule
\end{tabular}
}
\end{table*}

{\color{black}
For same-problem strong scaling, $T_N$ denotes the measured time for one
complete PIC step on $N$ GPUs; smaller $T_N$ is better.  Three runs are made
at every GPU count, and $T_N$ is the median of their mean step times.  The
smallest configuration that fits in memory is denoted by $N_{\min}$ (five
GPUs here), and it is used as the reference rather than an unavailable
single-GPU run.  We report
\begin{equation}
 \begin{aligned}
 S(N)&=\frac{T_{N_{\min}}}{T_N}, &
 E(N)&=\frac{S(N)}{N/N_{\min}},\\
 R(N)&=\frac{8T_8}{NT_N}.&&
 \end{aligned}
 \label{eq:evaluation_scaling_metrics}
\end{equation}
The speedup $S(N)$ answers ``how many times faster is $N$ GPUs than the
smallest feasible run?''  The efficiency $E(N)$ divides that speedup by the
ideal increase in GPU count.  For example, increasing from five to eight GPUs
would ideally accelerate the calculation by $8/5=1.6$; the measured speedup
of 1.469 therefore gives $E(8)=1.469/1.6=91.81\%$.  The resource ratio $R(N)$
compares total GPU-seconds per step with the eight-GPU run: $R>1$ means that
the $N$-GPU configuration consumes fewer GPU-seconds, even if it takes longer
in wall time.

Particle throughput counts one update whenever an electron is advanced and
also counts the ion updates performed at the configured subcycle.  Poisson
throughput is the logical mesh-cell count divided by the measured Poisson
time.  Multi-GPU stage time is the maximum over device workers, because the
next stage cannot start until the slowest worker finishes; communication and
load imbalance therefore remain visible.  HDF5 output, checkpoints, and final
particle snapshots were disabled during formal scaling, but are included in
the separate production planning rate where stated.

For the resolution comparison, the fine $r$--$z$ plane is restricted to the
nested coarse plane by averaging each $2\times2$ fine block.  With cylindrical
cell weights $w_i\propto r_i\Delta r_i\Delta z_i$, the reported errors are
\begin{equation}
 \epsilon_2=\left[
 \frac{\sum_i w_i(f_i-c_i)^2}{\sum_i w_i c_i^2}
 \right]^{1/2},\qquad
 \epsilon_\infty=\frac{\max_i|f_i-c_i|}{\max_i|c_i|},
 \label{eq:evaluation_field_metrics}
\end{equation}
together with the Pearson correlation $C(f,c)$.  The nearest cell-centered
azimuthal planes differ by approximately $25~\mu$m in arc position.  This is
small compared with the intended morphology and statistical comparison, but
the test is not a full three-dimensional norm.
}

\subsection{Poisson validation and mode scaling}
\label{sec:evaluation_poisson}

The first validation compares a 64-core CPU solve with a single RTX 5090 on
two four-patch Hall configurations.  Both backends required 24 iterations.
For $6{,}750{,}208$ active cells, the median CPU and GPU solve times were
1.43445 and 0.249218~s; for $54{,}001{,}664$ active cells they were 4.24973
and 0.884463~s.  The corresponding GPU speedups were $5.756\times$ and
$4.805\times$.  The complete CPU/GPU fields agreed to relative $L_2$ errors
of $4.64\times10^{-16}$ and $6.83\times10^{-16}$, with relative
$L_\infty$ errors below $2.22\times10^{-15}$.  Thus the acceleration is not
obtained by changing the discrete operator or convergence target.

\begin{figure*}[t]
\centering
\includegraphics[width=\textwidth]{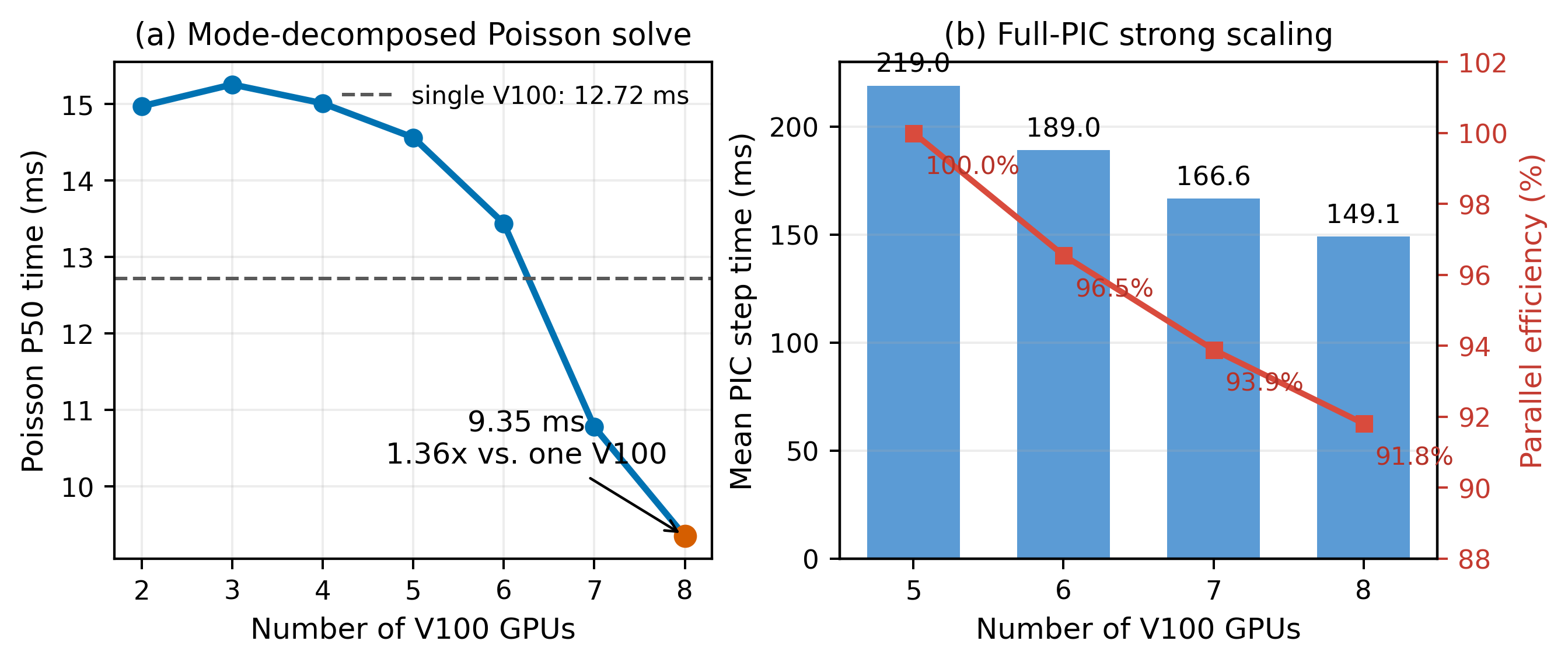}
\caption{\rev{Multi-V100 scaling, redrawn with text close to the manuscript
font size.  (a) End-to-end mode-decomposed Poisson P50 time
for all 65 physical modes; the eight-GPU point is approximately $1.36\times$
faster than one V100.  (b) Same-problem full-PIC step time and parallel
efficiency, using the five-GPU configuration as the smallest feasible
reference.}}
\label{fig:poisson_pic_scaling}
\end{figure*}

The multi-GPU test used the $512\times128\times400$ mesh and all 65 unique
real-to-complex Fourier modes.  Figure~\ref{fig:poisson_pic_scaling}(a) shows
the median end-to-end Poisson time for two through eight V100s.  The P50 times
were 14.971, 15.255, 15.009, 14.562, 13.439, 10.778, and 9.350~ms,
respectively.  The eight-GPU result was approximately $1.36\times$ faster
than the measured single-V100 reference.  All 21 runs passed: the maximum
relative $L_2$ and $L_\infty$ differences from the single-device solution
were $3.26\times10^{-15}$ and $8.17\times10^{-15}$, and the largest residual
was $1.17\times10^{-13}$.

Two to six GPUs do not outperform the single-device solver.  With only 65
independent modes, the reduction in local GMG work is initially smaller than
the fixed R2C, packing, reduce-scatter, all-gather, unpacking, and C2R costs.
Seven and eight devices provide enough aggregate GMG throughput and sufficiently
favorable NVLink paths to overcome that boundary cost.  This result supports
the communication placement in Section~\ref{sec:mode_gmg}: communication is
absent from a V-cycle, but the two spectral redistributions still set the
minimum profitable mode batch.

\subsection{Full-PIC strong scaling and layout invariance}
\label{sec:evaluation_pic_scaling}

The formal strong-scaling problem held the global mesh, particle state, source
models, and $707{,}788{,}800$ particles per species fixed.  Each GPU count
first passed a 40-step full-population memory and numerical precheck.  Three
independent 1200-step rounds were then executed; the first 200 steps were
discarded and the following 1000 steps were summarized.  The coefficient of
variation of the three per-round means was below 0.04\% for every feasible
configuration.

\begin{table*}[t]
\centering
\caption{Full-PIC strong scaling on V100.  Speedup and efficiency use five
GPUs as the smallest memory-feasible reference.  $R>1$ denotes lower GPU-second
cost than the eight-GPU run.}
\label{tab:pic_strong_scaling}
\small
\begin{tabular}{rrrrrrrr}
\toprule
GPUs & $T_N$ (s/step) & step/s & $S(N)$ & $E(N)$ (\%) & $R(N)$
 & minimum free memory (GiB) & maximum load ratio \\
\midrule
5 & 0.219028 & 4.566 & 1.000 & 100.00 & 1.089 & 5.697 & 1.0432 \\
6 & 0.189041 & 5.290 & 1.159 & 96.55  & 1.052 & 9.560 & 1.0442 \\
7 & 0.166635 & 6.001 & 1.314 & 93.89  & 1.023 & 12.228 & 1.0479 \\
8 & 0.149108 & 6.707 & 1.469 & 91.81  & 1.000 & 14.164 & 1.0565 \\
\bottomrule
\end{tabular}
\end{table*}

As shown in Table~\ref{tab:pic_strong_scaling} and
Fig.~\ref{fig:poisson_pic_scaling}(b), eight GPUs reduce wall time by
$1.469\times$ relative to five at 91.81\% parallel efficiency.  Five GPUs
use approximately 8.9\% fewer GPU-seconds but take 46.9\% longer per step.
Every feasible run retained at least 4~GiB free memory, a load ratio below
1.10, and a Poisson residual below $10^{-6}$.  Four GPUs were rejected during
initialization by a CUDA allocation failure while constructing the NCCL
reduce-scatter path for the production particle capacity.  It is therefore a
memory-infeasible configuration, not a numerical or topology failure.

Changing the number of particle slabs must not change the solution.  An
additional comparison between eight and five GPUs used 800,000 particles per
species with a common macro-particle weight.  In the 20-step source-free
comparison, all 18 field and
moment variables were finite and the largest relative $L_2$ and $L_\infty$
differences were $1.71\times10^{-10}$ and $1.49\times10^{-8}$.  In the
200-step source-enabled comparison, both layouts produced exactly the same
active populations, cathode count, ionization count, and absorbed-particle
counts.  Charge, energy, and residual diagnostics agreed within a relative
tolerance of $5\times10^{-6}$.  These tests isolate decomposition effects from
the resolution study and verify the ownership invariants of
Section~\ref{sec:particle_spatial}.

\subsection{Cylindrical Hall thruster plasma production study}
\label{sec:evaluation_production}

The production study compares two independently completed realizations of the
same cylindrical Hall thruster plasma model defined in
Section~\ref{sec:production_model}.  Table~\ref{tab:production_config} lists
the controlled and deliberately changed quantities.  The physical domain,
four patch boundaries, applied potential and magnetic field, source model and
rate, source temperatures, random seeds, and 50-ps ion-advance interval are
unchanged.  Refinement doubles the number of cells in every coordinate and
halves the electron time step.  It also changes particle sampling from 36 to
16 particles per active cell, so this is a practical resolution comparison
rather than a one-parameter convergence proof.

\revgreen{The principal numerical purpose of this comparison is to determine
whether the coarse spacing of approximately $100~\mu$m, which can exceed the
local electron Debye length, preserves the macroscopic solution obtained with
the approximately $50~\mu$m mesh.  The comparison therefore evaluates both
the local ratio $\Delta_{\max}/\lambda_D$ and the resulting field, density,
temperature, and transport structures.}

\begin{table}[t]
\centering
\caption{\rev{Cylindrical Hall thruster plasma production configurations.}}
\label{tab:production_config}
\small
\resizebox{\columnwidth}{!}{%
\begin{tabular}{lll}
\toprule
Quantity & RTX 5090 coarse & $8\times$V100 fine \\
\midrule
Grid $(N_r,N_\alpha,N_z)$ & $256\times64\times200$ & $512\times128\times400$ \\
Logical cells & $3{,}276{,}800$ & $26{,}214{,}400$ \\
Approximate cell scale & $100~\mu$m & $50~\mu$m \\
Electron time step & 5~ps & 2.5~ps \\
Ion subcycle & 10 & 20 \\
Initial particles/species & $88{,}473{,}600$ & $314{,}572{,}800$ \\
Particles/active cell/species & 36 & 16 \\
Macro-particle weight & 1388.889 & 390.625 \\
Poisson tolerance & $10^{-6}$ & $10^{-6}$ \\
Common 8-$\mu$s state & step 1,600,000 & step 3,200,000 \\
\bottomrule
\end{tabular}}
\end{table}

\begin{figure*}[t]
\centering
\includegraphics[width=\textwidth]{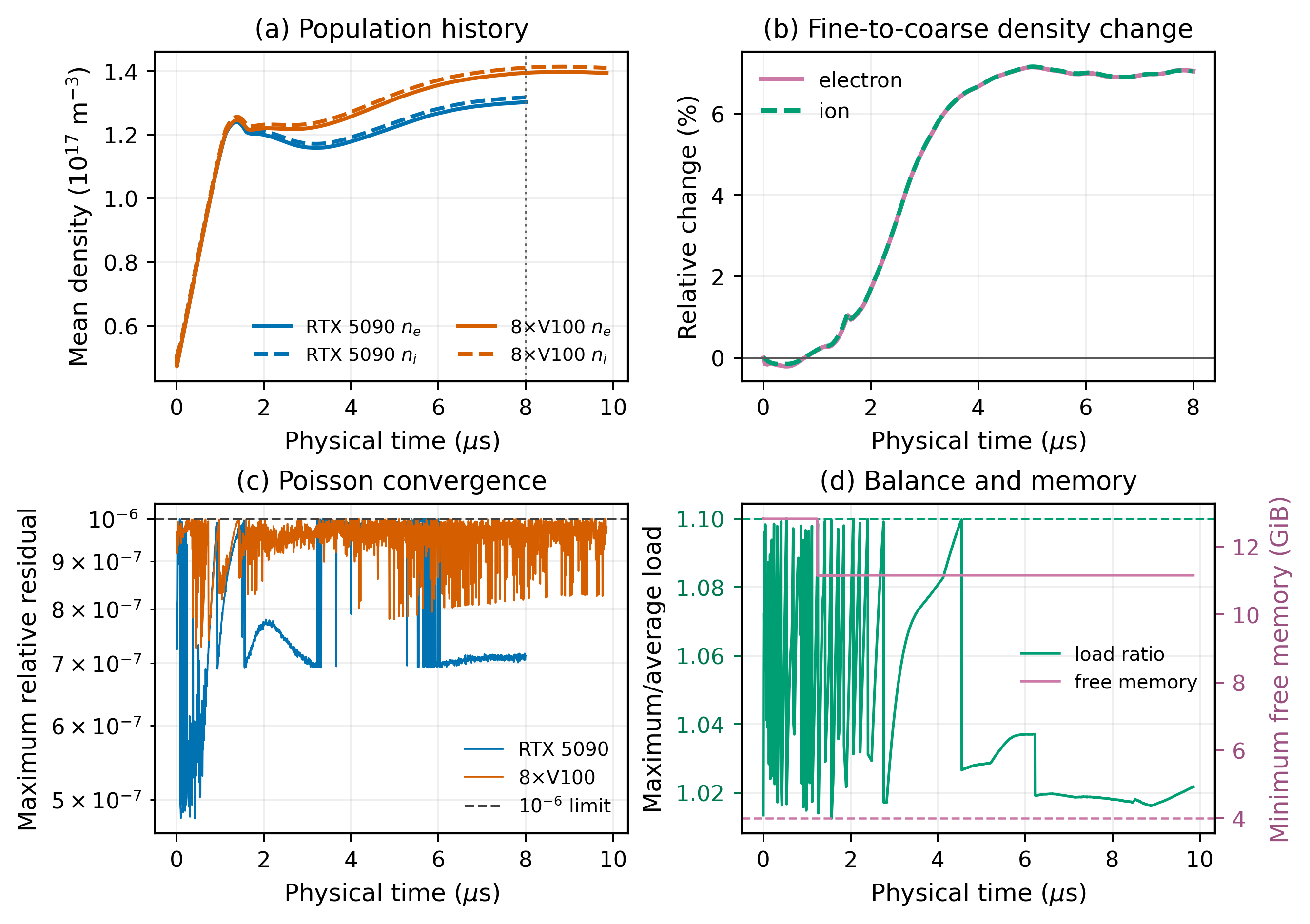}
\caption{\rev{Cylindrical Hall thruster plasma population and numerical
health, redrawn with larger labels.  (a) Coarse- and fine-grid
volume-averaged electron and ion densities.  (b) Fine-to-coarse density
change.  (c) Maximum Poisson residual.  (d) V100 particle-load ratio and
minimum free device memory.  Dashed lines denote the residual, load, and
memory acceptance thresholds.}}
\label{fig:population_health}
\end{figure*}

\paragraph{Population evolution and numerical health.}
Figure~\ref{fig:population_health} compares volume-averaged densities and the
archived solver-health diagnostics.  The two calculations reproduce the same
initial rise, transient decrease, and subsequent growth.  At 8~$\mu$s, the
fine-grid electron and ion densities are respectively 7.052\% and 7.046\%
higher than the coarse values.  The longer fine-grid trajectory reaches its
electron and ion population maxima near 8.800 and 8.825~$\mu$s and decreases
slightly by 9.855~$\mu$s, providing evidence for the onset of saturation
rather than indefinite growth.

The maximum archived Poisson residuals are $9.99957\times10^{-7}$ for the
RTX 5090 and $9.99988\times10^{-7}$ for the V100 calculation.  Neither archive
contains a nonfinite field, an invalid active particle, or a numerically
isolated particle.  Dynamic cut adjustment keeps the V100 load ratio below
1.10; at least 11.15~GiB remains free on every device.  The V100 trajectory
ended through a controlled signal after publishing a complete distributed
checkpoint, rather than through a numerical failure.

\begin{figure*}[t]
\centering
\includegraphics[width=\textwidth]{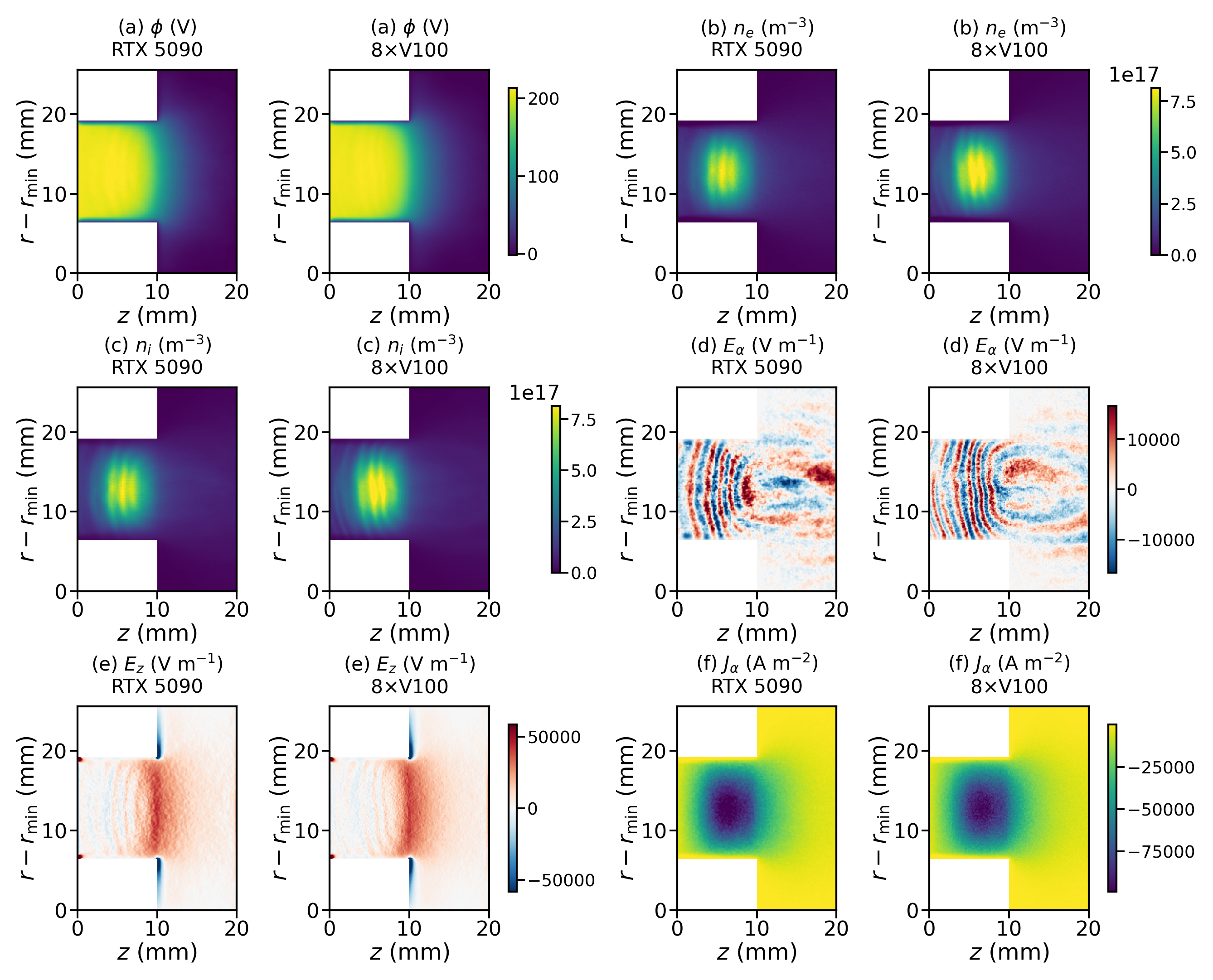}
\caption{\rev{Field comparison at 8~$\mu$s using larger panels and common
color limits within every coarse--fine pair.  The six pairs show $\phi$,
$n_e$, $n_i$, $E_\alpha$, $E_z$, and $J_\alpha$.  The left member of each
pair is the RTX~5090 coarse field and the right member is the eight-V100 field
restricted to the coarse grid.  Difference maps, $\rho$, and $J_z$ are
omitted; numerical differences are reported in
Table~\ref{tab:field_restriction_metrics}.}}
\label{fig:field_resolution_comparison}
\end{figure*}

\begin{figure*}[t]
\centering
\includegraphics[width=\textwidth]{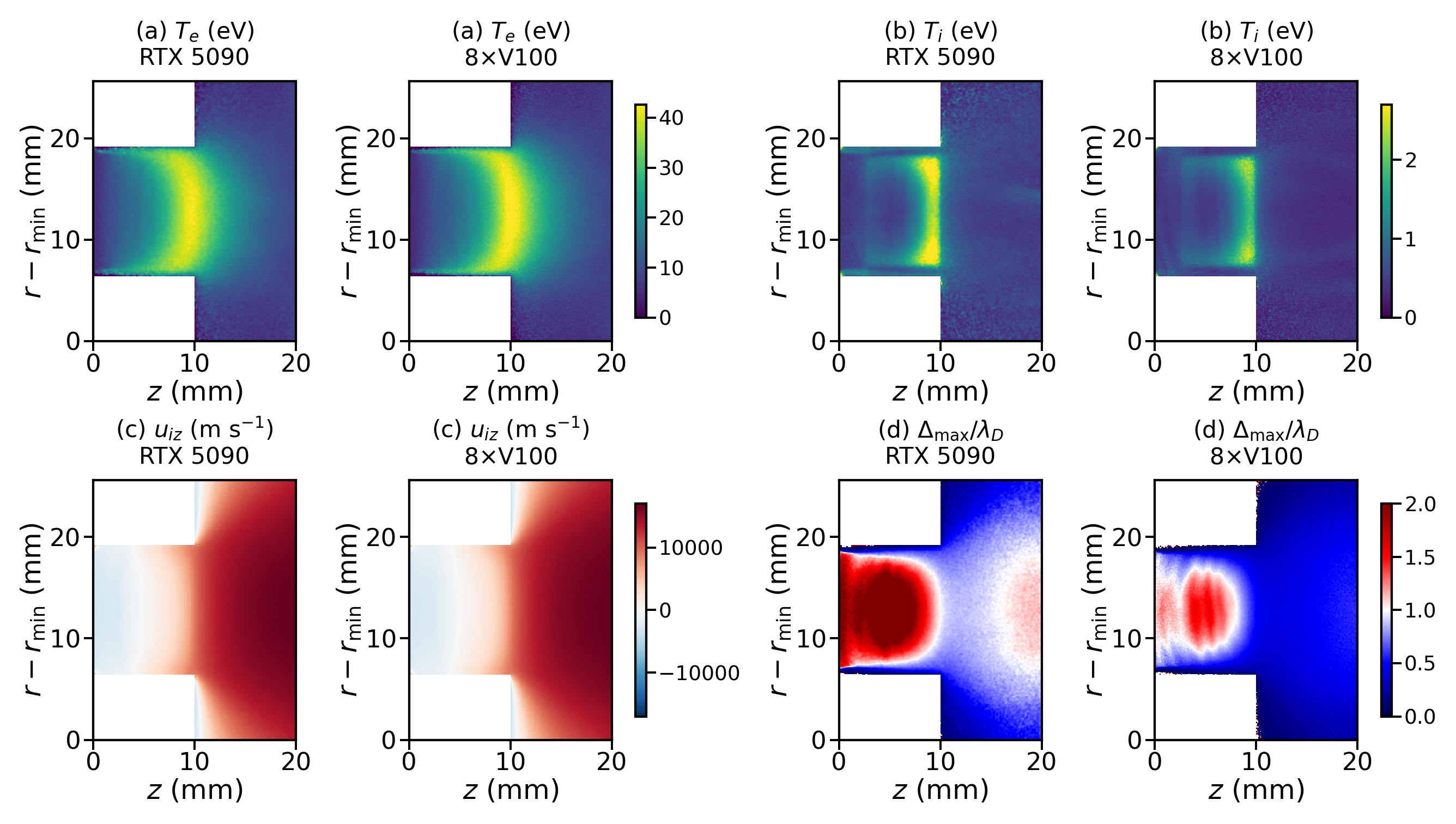}
\caption{\revgreen{Temperature, ion velocity, and Debye-length resolution at
8~$\mu$s, arranged as two rows of four panels.  Each adjacent pair compares
the RTX~5090 coarse result with the eight-V100 fine result: (a) $T_e$,
(b) $T_i$, (c) $u_{i,z}$, and (d) $\Delta_{\max}/\lambda_D$.  The field
quantities use the fine solution restricted to the coarse mesh and one color
scale per pair.  Panel (d) uses each native mesh, a fixed range of 0--2, and
the seismic diverging map with white at the nominal limit of one.}}
\label{fig:temperature_transport_comparison}
\end{figure*}

\paragraph{Restricted field comparison.}
{\color{black}
Figures~\ref{fig:field_resolution_comparison} and
\ref{fig:temperature_transport_comparison} place the RTX~5090 coarse field
beside the eight-V100 field restricted to the same coarse $z$--$r$ mesh at
8~$\mu$s.  Difference maps are omitted to keep each physical field readable;
the magnitude of the coarse--fine discrepancy is instead summarized by
$\epsilon_2$ and the spatial correlation $C$ in
Table~\ref{tab:field_restriction_metrics}.  Potential is nearly resolution
independent ($\epsilon_2=0.028$, $C=0.999$).  The electron and ion densities
retain the same channel--plume envelope, while $E_z$, $J_\alpha$, $T_e$,
$T_i$, and $u_{i,z}$ remain strongly correlated.  The instantaneous
$E_\alpha$ panels occupy the same wave-bearing region but do not coincide
pointwise in phase, which is reflected by the low value of $C$ rather than by
a visually compressed difference panel.
}

\paragraph{Debye-length resolution.}
\revgreen{For electron temperature $T_e$ expressed in electron-volts, the
local electron Debye length and the largest physical cell dimension are
defined as
\[
 \lambda_D=\left(\frac{\varepsilon_0T_e}{e n_e}\right)^{1/2},
 \qquad
 \Delta_{\max}=\max(\Delta r,\,r\Delta\alpha,\,\Delta z).
\]
The usual electrostatic PIC resolution guideline is satisfied when this
ratio does not exceed unity.
Figure~\ref{fig:temperature_transport_comparison}(d) plots this ratio on the
native coarse and fine meshes.  Both panels use the fixed interval 0--2 with
the seismic diverging map centered at one; white therefore marks the nominal
limit and values above two saturate at the red endpoint.  At 8~$\mu$s, the
cylindrically weighted median ratio decreases from 0.880 on the coarse mesh
to 0.433 on the fine mesh.  The cylindrically weighted fraction of the active
$z$--$r$ section satisfying the guideline increases from 64.9\% to 86.4\%.
\revblue{Thus, the coarse mesh satisfies the guideline over 64.9\% of the
weighted active section but not over the remaining 35.1\%; it is partially,
rather than wholly, under-resolved.  The fine mesh substantially reduces this
under-resolved fraction, although it also does not satisfy the criterion in
every cell.  In an instability-resolving three-dimensional Hall thruster PIC
calculation of this scale, enforcing $\Delta_{\max}/\lambda_D<1$ in every
active cell throughout a long evolution would require substantially more mesh
cells and particles in all three directions and is exceptionally demanding
computationally.  \revblue{Hall-thruster discharges are also open particle
systems: energetic electrons can be lost at material surfaces or through the
plume outflow after a finite residence time.  This turnover provides a loss
channel for electrons affected by finite-grid heating and can limit the
indefinite accumulation of that numerical error, although it does not remove
the local resolution requirement.  In our previous three-dimensional PIC
grid-sensitivity study, a baseline mesh that was not Debye-resolved everywhere
and a refined mesh nevertheless recovered the same large-scale near-wall
anomalous-transport topology, with refinement primarily adding local spatial
detail~\cite{LiuZhaoZhao2026NearWall}.}}
Nevertheless, its potential, density envelope, axial field, azimuthal current,
temperature, and ion-velocity structures remain consistent with the refined
calculation.  For this operating condition, the coarse mesh therefore
reproduces the large-scale evolution and statistical transport even where the
strict local grid criterion is not met.  This agreement does not establish
Debye-scale accuracy in sheaths or for phase-sensitive high-frequency
fluctuations.  \revblue{Because the present study is focused on the multi-GPU
algorithm, numerical equivalence, and parallel performance rather than on a
new convergence study of Hall-thruster physics, the Debye-ratio analysis is
used here to delimit the physical interpretation, not to claim Debye-scale
accuracy.}}

\begin{table*}[t]
\centering
\caption{\revgreen{Quantitative field-restriction and matched-window performance
results.  In (a), $C$ is the Pearson spatial correlation and the listed
variables correspond to Figs.~\ref{fig:field_resolution_comparison} and
\ref{fig:temperature_transport_comparison}(a)--(c); the Debye-length metric in
panel (d) is summarized in the text.  In (b), ratios are fine/coarse except
for throughput, where they denote aggregate speed relative to the RTX 5090.}}
\label{tab:field_restriction_metrics}
\label{tab:production_performance}
\small
{\color{black}
\begin{minipage}[t]{0.42\textwidth}
\centering
\textbf{(a) Cylindrically weighted field comparison}\par\smallskip
\begin{tabular}{lrr}
\toprule
Variable & $\epsilon_2$ & $C$ \\
\midrule
$\phi$ & 0.028 & 0.999 \\
$n_e$ & 0.122 & 0.996 \\
$n_i$ & 0.134 & 0.994 \\
$E_\alpha$ & 1.404 & $-0.042$ \\
$E_z$ & 0.212 & 0.975 \\
$J_\alpha$ & 0.085 & 0.995 \\
$T_e$ & 0.105 & 0.982 \\
$T_i$ & 0.257 & 0.955 \\
$u_{i,z}$ & 0.019 & 1.000 \\
\bottomrule
\end{tabular}
\end{minipage}\hfill
\begin{minipage}[t]{0.54\textwidth}
\centering
\textbf{(b) Production performance near 1.235~$\mu$s}\par\smallskip
\begin{tabular}{lrrr}
\toprule
Metric & RTX 5090 & $8\times$V100 & Ratio \\
\midrule
Step time (s) & 0.15584 & 0.17737 & 1.138 \\
Grid cells (million) & 3.277 & 26.214 & 8.000 \\
Particle updates (billion/s) & 1.536 & 4.593 & 2.991 \\
Poisson cells (million/s) & 269.8 & 815.1 & 3.021 \\
Wall hours/simulated $\mu$s & 8.66 & 19.71 & 2.276 \\
GPU-hours/simulated $\mu$s & 8.66 & 157.68 & 18.21 \\
\bottomrule
\end{tabular}
\end{minipage}
}
\end{table*}

\begin{figure*}[t]
\centering
\includegraphics[width=0.94\textwidth]{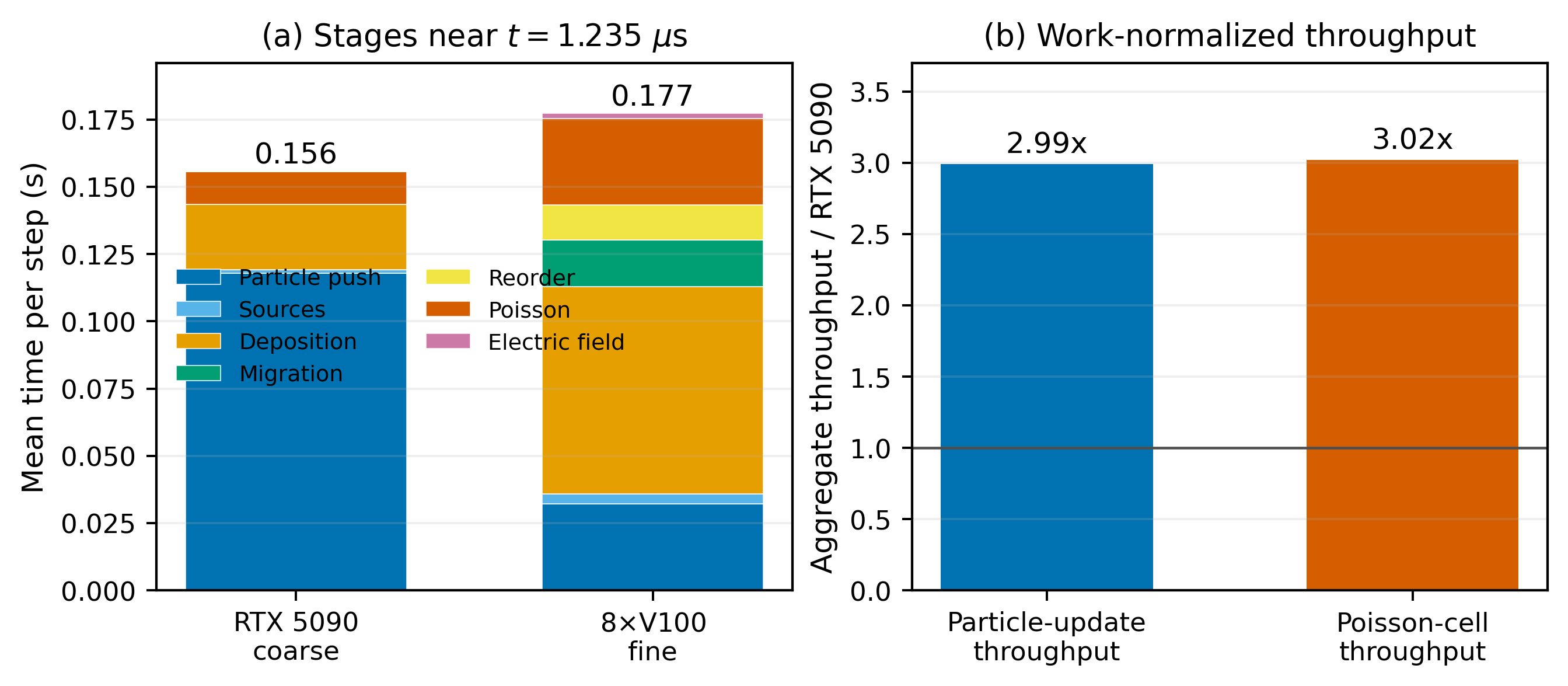}
\caption{\rev{Matched production window near $t=1.235~\mu$s, redrawn with
larger labels.  Left: mean stage
time per numerical step.  Right: work-normalized particle-update and
Poisson-cell throughput relative to the single RTX 5090.  These aggregate
throughput ratios are not energy-efficiency measurements.}}
\label{fig:performance_breakdown}
\end{figure*}

\paragraph{Cross-platform production throughput.}
Performance is compared near $t=1.235~\mu$s, before the later particle-count
divergence.  The V100 value is the mean of 1000 contiguous steps and contains
exactly 50 ion-subcycle steps; the RTX 5090 value is the corresponding archived
diagnostic.  Table~\ref{tab:production_performance} and
Fig.~\ref{fig:performance_breakdown} show that one fine step takes 0.17737~s,
only 13.8\% longer than a coarse step, despite processing eight times as many
grid cells and approximately 3.57 times as many active particles at that
window.  Aggregate particle-update and Poisson-cell throughput are
$2.99\times$ and $3.02\times$ the RTX 5090 values.

Refinement halves the electron time step, so twice as many steps are required
per unit physical time.  The fine trajectory therefore costs $2.28\times$
more wall time and $18.2\times$ more aggregate GPU-hours.  The stage breakdown
also reveals a different bottleneck: particle push accounts for 75.6\% of the
RTX 5090 step, whereas V100 deposition is the largest stage at 43.5\%.
Poisson and particle push each contribute approximately 18\%, and migration
plus periodic reorder contribute 17.1\%.  Continued multi-GPU optimization
should therefore prioritize cell-local deposition and lower-cost particle
migration/reordering in addition to mode-solver tuning.

\subsection{Interpretation and limitations}
\label{sec:evaluation_limits}

The experiments establish three distinct claims.  First, the full-spectrum
batched operator reproduces reference fields to double-precision roundoff and
retains all physical modes.  Second, the identical full-PIC problem
strong-scales from five to eight V100s at 91.81\% efficiency while preserving
the solution across particle layouts.  Third, the dual-decomposition code can
sustain a restartable, long-duration calculation containing hundreds of
millions of particles and an eight-times-larger field mesh.

The RTX 5090/V100 comparison should not be interpreted as a hardware speedup:
both hardware generation and numerical resolution change.  It demonstrates
aggregate throughput sufficient to make the refined problem practical.  The
field comparison likewise supports resolution-consistent macroscopic physics
and statistical transport, but not a formal convergence order because the
particles per cell also change and only two resolutions are available.
Instantaneous high-frequency quantities require spectra, phase-speed analysis,
longer time averages, and multiple random seeds before their uncertainty can
be separated from resolution sensitivity.  Finally, no GPU power data were
recorded; GPU-hours quantify resource use, not energy efficiency.

\section{Conclusions}
\label{sec:conclusions}

The principal result is that a complete PIC step need not force its field and
particle stages to share one parallel decomposition.  Exact, full-spectrum
Fourier diagonalization gives the globally coupled Poisson solve a
complete-mode ownership layout, whereas the particle stage uses spatial
ownership.  Reconstructing the complete field on every GPU connects these
layouts without field halos or redundant Poisson solves.  In the field path,
collectives occur only before and after the batched matrix-free multigrid solve,
so the V-cycle itself contains no inter-device communication.  In the particle
path, direct peer migration, cell-major reorder, warp-aggregated CIC, and
species- and capacity-aware cuts jointly recover locality and control evolving
population imbalance.  Their integration, while preserving all discrete modes,
the represented particle set, and the eight-corner CIC rule, is the main
algorithmic contribution.

The solver and end-to-end evidence support this contribution.  CPU and
single-GPU Poisson fields agree to relative $L_2$ errors below
$6.9\times10^{-16}$.  Eight V100 GPUs solve all 65 physical modes in 9.350~ms
and agree with the single-device result to below $8.2\times10^{-15}$.  The
identical $707{,}788{,}800$-particle-per-species PIC problem strong-scales from
five to eight V100 GPUs with 91.81\% efficiency, reaches 0.149108~s per step,
and satisfies the residual, memory, and load-balance gates.  The long
cylindrical Hall thruster plasma calculation further demonstrates checkpointed
operation with hundreds of millions of particles.  Relative to a single
RTX~5090 coarse discretization, the eight-V100 refined calculation provides
approximately threefold aggregate particle-update and Poisson-cell throughput.
Thus, the contribution is an application-level decomposition strategy, not
merely acceleration of an isolated CUDA kernel.

The cross-platform production comparison remains deliberately limited:
hardware, mesh resolution, particle sampling, and time step all change, so it
is neither a hardware speedup nor a formal grid-convergence study.  The fields
support agreement of macroscopic structure and averaged transport, whereas
phase-sensitive fluctuations require spectra, longer statistical windows, and
multiple random realizations.  Power was not recorded, so GPU-hours are not an
energy-efficiency metric.  Future work will reduce the remaining deposition
and migration/reordering costs, extend the protected-face hierarchy to more
general materials and boundaries, and quantify fluctuation uncertainty without
weakening the full-spectrum formulation.

\section*{Data availability}
The source code, input configurations, and supporting numerical data are available from the corresponding author upon reasonable request.

\section*{CRediT authorship contribution statement}
\textbf{Yinjian Zhao:} Conceptualization, Methodology, Software, Validation,
Formal analysis, Investigation, Data curation, Visualization, Writing --
original draft, Writing -- review \& editing, Project administration, Funding
acquisition, Resources, Supervision. \textbf{Xi Chen:} Conceptualization,
Writing -- review \& editing. \textbf{Yingjie Chen:} Conceptualization,
Writing -- review \& editing.

\section*{Declaration of competing interest}
The authors declare that they have no known competing financial interests or
personal relationships that could have appeared to influence the work reported
in this paper.

\section*{Acknowledgements}
\revblue{The authors acknowledge support from the National Natural Science Foundation of China (Grant Nos. 52472403).}

\appendix
\section{From the mode equation to the matrix-free cell stencil}
\label{app:fv_shifted_operator}

This appendix gives the intermediate steps between the differential mode equation in Eq.~\eqref{eq:mode_poisson} and the matrix-free operation in Eq.~\eqref{eq:matrix_free_mode_operator}. The derivation is first written for an interior cell with four active neighbors, followed by the corresponding treatment of a Dirichlet face and a numerical stencil example.

\subsection{Cell integration and face fluxes}

Consider the active cell $q=(i,k)$, where $i$ and $k$ label radial and axial cells. The cell is bounded by radial faces $r_{i-1/2}$ and $r_{i+1/2}$ and axial faces $z_{k-1/2}$ and $z_{k+1/2}$; half-integer subscripts label cell faces. Its center coordinates and widths are
\begin{equation}
\begin{aligned}
r_i&=\tfrac12\left(r_{i-1/2}+r_{i+1/2}\right),
&\Delta r_i&=r_{i+1/2}-r_{i-1/2},\\
z_k&=\tfrac12\left(z_{k-1/2}+z_{k+1/2}\right),
&\Delta z_k&=z_{k+1/2}-z_{k-1/2}.
\end{aligned}
\label{eq:cell_geometry_appendix}
\end{equation}
Let $W=(i-1,k)$, $E=(i+1,k)$, $S=(i,k-1)$, and $N=(i,k+1)$ denote the west, east, south, and north neighbors of $q$, respectively. For Fourier mode $m$, write $u_{q,m}=\widehat{\phi}_m(r_i,z_k)$ for the cell-centered potential and use $u_{W,m}$, $u_{E,m}$, $u_{S,m}$, and $u_{N,m}$ for the four neighboring values.

The face conductances are defined by
\begin{equation}
\begin{aligned}
G_{r,-}&=\frac{r_{i-1/2}\Delta z_k}{r_i-r_{i-1}},
&G_{r,+}&=\frac{r_{i+1/2}\Delta z_k}{r_{i+1}-r_i},\\
G_{z,-}&=\frac{r_i\Delta r_i}{z_k-z_{k-1}},
&G_{z,+}&=\frac{r_i\Delta r_i}{z_{k+1}-z_k}.
\end{aligned}
\label{eq:interior_face_conductances}
\end{equation}
Here, $G_{r,-}$ and $G_{r,+}$ are the lower- and upper-radial conductances, while $G_{z,-}$ and $G_{z,+}$ are their axial counterparts. The coordinates $r_{i-1}$, $r_{i+1}$, $z_{k-1}$, and $z_{k+1}$ are neighboring cell centers.

Integrating the first term of Eq.~\eqref{eq:mode_poisson} over the cell rectangle and applying the fundamental theorem of calculus in $r$ gives
\begin{align}
&-\int_{z_{k-1/2}}^{z_{k+1/2}}
  \int_{r_{i-1/2}}^{r_{i+1/2}}
  \frac{\partial}{\partial r}
  \left(r\frac{\partial\widehat{\phi}_m}{\partial r}\right)
  \,\mathrm{d}r\,\mathrm{d}z \notag\\
&\quad=-\int_{z_{k-1/2}}^{z_{k+1/2}}
  \left[r\frac{\partial\widehat{\phi}_m}{\partial r}\right]_{r_{i-1/2}}^{r_{i+1/2}}
  \,\mathrm{d}z \notag\\
&\quad\approx
G_{r,-}(u_{q,m}-u_{W,m})
+G_{r,+}(u_{q,m}-u_{E,m}).
\label{eq:radial_flux_balance}
\end{align}
The last line uses a centered difference between the two cell centers adjacent to each face. The sign is worth noting: the negative divergence produces a positive coefficient for the current cell and a negative coefficient for each neighbor.

For the axial term, integration in $z$ converts the second derivative into its two face gradients. In addition,
$\int_{r_{i-1/2}}^{r_{i+1/2}}r\,\mathrm{d}r=r_i\Delta r_i$ because $r_i$ is the midpoint of the radial faces. Therefore,
\begin{align}
&-\int_{r_{i-1/2}}^{r_{i+1/2}}
  \int_{z_{k-1/2}}^{z_{k+1/2}}
  r\frac{\partial^2\widehat{\phi}_m}{\partial z^2}
  \,\mathrm{d}z\,\mathrm{d}r \notag\\
&\quad\approx
G_{z,-}(u_{q,m}-u_{S,m})
+G_{z,+}(u_{q,m}-u_{N,m}).
\label{eq:axial_flux_balance}
\end{align}

The angular shift and the charge term contain no derivatives in the $r$--$z$ plane. Midpoint quadrature consequently gives
\begin{align}
\int_{q}\frac{\lambda_m}{r}\widehat{\phi}_m
\,\mathrm{d}r\,\mathrm{d}z
&\approx \lambda_m\mu_q u_{q,m},
&\mu_q&=\frac{\Delta r_i\Delta z_k}{r_i},
\label{eq:integrated_mode_shift}\\
\int_{q}\frac{r}{\epszero}\widehat{\rho}_m
\,\mathrm{d}r\,\mathrm{d}z
&\approx s_q\widehat{\rho}_{q,m},
&s_q&=\frac{r_i\Delta r_i\Delta z_k}{\epszero}.
\label{eq:integrated_charge_rhs}
\end{align}
Here, $\lambda_m$ is the exact discrete angular eigenvalue from Eq.~\eqref{eq:discrete_azimuthal_eigenvalue}, $\mu_q$ is the cell coefficient of the mode-dependent shift, $\widehat{\rho}_{q,m}=\widehat{\rho}_m(r_i,z_k)$ is the cell-centered charge-density coefficient, and $s_q$ is its finite-volume scaling factor.

For an interior cell, define the mode-independent diagonal
\begin{equation}
D_q^{(0)}=G_{r,-}+G_{r,+}+G_{z,-}+G_{z,+}.
\label{eq:interior_base_diagonal}
\end{equation}
Adding Eqs.~\eqref{eq:radial_flux_balance}--\eqref{eq:integrated_charge_rhs} then produces the complete interior-cell balance
\begin{align}
&\left(D_q^{(0)}+\lambda_m\mu_q\right)u_{q,m}
-G_{r,-}u_{W,m}-G_{r,+}u_{E,m}\notag\\
&\quad-G_{z,-}u_{S,m}-G_{z,+}u_{N,m}
=s_q\widehat{\rho}_{q,m}.
\label{eq:explicit_interior_cell_balance}
\end{align}
If $\mathcal{N}(q)=\{W,E,S,N\}$ is the neighbor set and $G_{qq'}$ is the conductance connecting $q$ to neighbor $q'$, the left-hand side of Eq.~\eqref{eq:explicit_interior_cell_balance} is
\begin{equation}
(A_m u_m)_q
=\left(D_q^{(0)}+\lambda_m\mu_q\right)u_{q,m}
-\sum_{q'\in\mathcal{N}(q)}G_{qq'}u_{q',m},
\label{eq:appendix_matrix_row}
\end{equation}
which is Eq.~\eqref{eq:matrix_free_mode_operator}. Thus, one application of $A_m$ is simply an evaluation of the cell balance: multiply the current-cell value by its diagonal, subtract the four conductance-weighted neighbor values, and repeat this operation for every active cell. The sparse matrix need not be assembled or stored.

\subsection{Dirichlet faces}

If a face lies on a Dirichlet boundary instead of adjoining an active neighbor, its conductance uses the distance from the cell center to the boundary face:
\begin{equation}
\begin{aligned}
G^D_{r,-}&=\frac{r_{i-1/2}\Delta z_k}{r_i-r_{i-1/2}},
&G^D_{r,+}&=\frac{r_{i+1/2}\Delta z_k}{r_{i+1/2}-r_i},\\
G^D_{z,-}&=\frac{r_i\Delta r_i}{z_k-z_{k-1/2}},
&G^D_{z,+}&=\frac{r_i\Delta r_i}{z_{k+1/2}-z_k}.
\end{aligned}
\label{eq:dirichlet_face_conductances}
\end{equation}
The superscript $D$ denotes a Dirichlet face. Let $f$ label such a face and let $\widehat{\phi}_{f,m}^D$ be its prescribed mode-$m$ potential. Its flux contribution is $G_f^D(u_{q,m}-\widehat{\phi}_{f,m}^D)$. Consequently, $G_f^D$ is added to the diagonal, whereas the known term $G_f^D\widehat{\phi}_{f,m}^D$ is moved to the right-hand side.

Let $\mathcal{F}_D(q)$ be the set of Dirichlet faces of cell $q$. Including both active neighbors and Dirichlet faces gives
\begin{align}
D_q^{(0)}
&=\sum_{q'\in\mathcal{N}(q)}G_{qq'}
+\sum_{f\in\mathcal{F}_D(q)}G_f^D,
\label{eq:base_diagonal_with_boundary}\\
b_{q,m}^D
&=\sum_{f\in\mathcal{F}_D(q)}
G_f^D\widehat{\phi}_{f,m}^D,
\label{eq:boundary_rhs_appendix}
\end{align}
where $b_{q,m}^D$ is the boundary contribution to the algebraic right-hand side. For an azimuthally invariant boundary, $\widehat{\phi}_{f,m}^D=\phi_f^D\delta_{m0}$, where $\phi_f^D$ is the prescribed physical potential and $\delta_{m0}$ is the Kronecker delta. The full cell equation is therefore
\begin{equation}
(A_m u_m)_q
=s_q\widehat{\rho}_{q,m}+b_{q,m}^D.
\label{eq:full_finite_volume_row}
\end{equation}

\subsection{Worked five-point stencil example}

Consider an illustrative normalized cell with $\Delta r_i=\Delta z_k=1$, center radius $r_i=2$, radial faces $r_{i-1/2}=1.5$ and $r_{i+1/2}=2.5$, and unit center-to-center distances in all four directions. Assume first that all four neighbors are active. Equations~\eqref{eq:interior_face_conductances}, \eqref{eq:integrated_mode_shift}, and \eqref{eq:integrated_charge_rhs} give
\begin{equation}
\begin{aligned}
G_{r,-}&=1.5, & G_{r,+}&=2.5,\\
G_{z,-}&=2,   & G_{z,+}&=2,\\
\mu_q&=0.5,   & s_q&=\frac{2}{\epszero}.
\end{aligned}
\label{eq:worked_coefficients}
\end{equation}
Substitution into Eq.~\eqref{eq:explicit_interior_cell_balance} yields
\begin{align}
&(8+0.5\lambda_m)u_{q,m}
-1.5u_{W,m}-2.5u_{E,m}\notag\\
&\qquad-2u_{S,m}-2u_{N,m}
=\frac{2}{\epszero}\widehat{\rho}_{q,m}.
\label{eq:worked_interior_stencil}
\end{align}
For the local ordering $(q,W,E,S,N)$, the corresponding row of $A_m$ is
\begin{equation}
\begin{array}{c|ccccc}
\text{column} & q & W & E & S & N\\ \hline
\text{entry} & 8+0.5\lambda_m & -1.5 & -2.5 & -2 & -2
\end{array}.
\label{eq:worked_matrix_row}
\end{equation}
For the zero mode, $\lambda_0=0$, so the center entry is $8$ and the four neighbor entries are $-1.5$, $-2.5$, $-2$, and $-2$. Any nonzero mode changes only the center entry by $0.5\lambda_m$; all neighbor coefficients remain unchanged. In matrix notation, the number $0.5$ is the $q$th diagonal entry of $M$, making this row a concrete example of $A_m=A_0+\lambda_mM$.

As a boundary example, replace the east neighbor by a Dirichlet face at $r_{i+1/2}=2.5$. Equation~\eqref{eq:dirichlet_face_conductances} gives $G_{r,+}^D=2.5/(2.5-2)=5$. If the prescribed Fourier coefficient is $\widehat{\phi}_{E,m}^D$, the cell equation becomes
\begin{align}
&(10.5+0.5\lambda_m)u_{q,m}
-1.5u_{W,m}-2u_{S,m}-2u_{N,m}\notag\\
&\qquad=\frac{2}{\epszero}\widehat{\rho}_{q,m}
+5\widehat{\phi}_{E,m}^D.
\label{eq:worked_dirichlet_stencil}
\end{align}
The boundary conductance $5$ has been added to the diagonal, and the prescribed boundary value appears on the right-hand side. This example also shows why the matrix-free kernel needs only cell geometry, neighbor indices, boundary metadata, and the scalar $\lambda_m$: the off-diagonal topology and conductances are common to all Fourier modes.

\subsection{From the cell stencil to the shifted matrix family}

Equation~\eqref{eq:shifted_operator_family} follows by comparing the same cell row at mode $m$ and at the zero mode. Suppose the cross-section contains $Q$ active cells, and collect their potentials into
\begin{equation}
u_m=\left(u_{1,m},u_{2,m},\ldots,u_{Q,m}\right)^{\mathsf T},
\label{eq:appendix_mode_vector}
\end{equation}
where $Q$ is the number of packed cells and the superscript $\mathsf T$ denotes transpose. For an arbitrary cell $q$, Eq.~\eqref{eq:appendix_matrix_row} gives
\begin{align}
(A_m u_m)_q
&=D_q^{(0)}u_{q,m}
-\sum_{q'\in\mathcal{N}(q)}G_{qq'}u_{q',m}
+\lambda_m\mu_q u_{q,m},
\label{eq:mode_row_split}\\
(A_0 u_m)_q
&=D_q^{(0)}u_{q,m}
-\sum_{q'\in\mathcal{N}(q)}G_{qq'}u_{q',m},
\label{eq:zero_mode_row_appendix}
\end{align}
because $\lambda_0=0$. The vector $u_m$ is used in both lines only to compare the two operators acting on the same arbitrary input. Subtracting Eq.~\eqref{eq:zero_mode_row_appendix} from Eq.~\eqref{eq:mode_row_split} leaves
\begin{equation}
\bigl[(A_m-A_0)u_m\bigr]_q
=\lambda_m\mu_q u_{q,m}.
\label{eq:row_operator_difference}
\end{equation}

Define the diagonal matrix
\begin{equation}
M=\operatorname{diag}(\mu_1,\mu_2,\ldots,\mu_Q).
\label{eq:appendix_mass_matrix}
\end{equation}
By definition, the $q$th component of $Mu_m$ is $(Mu_m)_q=\mu_q u_{q,m}$. Equation~\eqref{eq:row_operator_difference} therefore holds for every cell and can be written as
\begin{equation}
(A_m-A_0)u_m=\lambda_mMu_m.
\label{eq:vector_operator_difference}
\end{equation}
Since this identity is valid for an arbitrary vector $u_m$, the two linear operators themselves must satisfy
\begin{equation}
A_m-A_0=\lambda_mM,
\qquad\text{or equivalently}\qquad
A_m=A_0+\lambda_mM,
\label{eq:appendix_shifted_operator_proof}
\end{equation}
which is Eq.~\eqref{eq:shifted_operator_family}.

A two-cell example makes the matrix addition explicit. Let cells 1 and 2 share a face of conductance $2$. Suppose their complete mode-independent diagonal coefficients, including their other faces, are $8$ and $9$, and let their shift coefficients be $\mu_1=0.5$ and $\mu_2=0.4$. The zero-mode operator and the diagonal shift matrix are
\begin{equation}
A_0=\begin{bmatrix}8&-2\\-2&9\end{bmatrix},
\qquad
M=\begin{bmatrix}0.5&0\\0&0.4\end{bmatrix}.
\label{eq:worked_two_cell_base_matrices}
\end{equation}
For mode $m$, scalar multiplication and matrix addition give
\begin{equation}
A_m=A_0+\lambda_mM
=\begin{bmatrix}
8+0.5\lambda_m&-2\\
-2&9+0.4\lambda_m
\end{bmatrix}.
\label{eq:worked_two_cell_shifted_matrix}
\end{equation}
For example, if $\lambda_m=3$ in the normalized units of this illustration, then
\begin{equation}
A_m=\begin{bmatrix}9.5&-2\\-2&10.2\end{bmatrix}.
\label{eq:worked_two_cell_numeric_matrix}
\end{equation}
The shared-face entries remain $-2$ for every mode; only the first diagonal changes from $8$ to $8+0.5\lambda_m$, and the second changes from $9$ to $9+0.4\lambda_m$. This is the algebraic reason that all modes can reuse the same neighbor graph, conductances, and multigrid transfer maps while the GPU kernel supplies only a mode-dependent diagonal shift.

\section{Worked examples for the spatial particle pipeline}
\label{app:particle_pipeline}

This appendix gives small numerical examples for the operations described in Section~\ref{sec:particle_spatial}. The examples are deliberately one-dimensional in their bookkeeping; the production kernels retain all three particle coordinates, all three velocity components, and the complete composite-domain cell state.

\subsection{Deposition and migration across one axial cut}

Consider \(P=3\) logical GPUs, \(N_z=10\) axial cells, and the cut vector
\[
(c_0,c_1,c_2,c_3)=(0,4,7,10).
\]
Equation~\eqref{eq:particle_z_owner} assigns cells \(0\)--\(3\) to logical GPU 0, cells \(4\)--\(6\) to logical GPU 1, and cells \(7\)--\(9\) to logical GPU 2. Suppose that a particle stored on GPU 0 starts in cell 3 and, after the push, has cached axial index \(i_z(p)=4\). Its old owner is \(g_{\mathrm{old}}(p)=0\), while its new owner is \(g(p)=1\); hence \(p\in\mathcal O_{0\rightarrow1}\).

The source device first deposits this particle at its new physical position. This is valid because GPU 0 has the complete field mesh and a complete deposition array, including cell 4. If \(\delta\rho_p\) denotes the eight-node CIC contribution of this particle and \(\rho_g^{-}\) denotes the array deposited on GPU \(g\) before migration, then
\[
\sum_{g=0}^{2}\rho_g^{-}
=\delta\rho_p+\sum_{q\ne p}\delta\rho_q,
\]
where \(q\) indexes every other active particle. Moving the record to GPU 1 afterward changes only the device that stores \(p\). If deposition were repeated conceptually after migration, the right-hand side would contain exactly the same contributions.

For the transfer, GPU 0 increments the destination count for GPU 1 and packs the complete record
\[
\mathcal Q_p=
\bigl(r_p,\alpha_p,z_p,\,
v_{r,p},v_{\alpha,p},v_{z,p},\,
i_r(p),i_\alpha(p),i_z(p),R(p)\bigr),
\]
where \(r_p\), \(\alpha_p\), and \(z_p\) are the cylindrical position coordinates; \(v_{r,p}\), \(v_{\alpha,p}\), and \(v_{z,p}\) are the corresponding velocity components; \(i_r(p)\), \(i_\alpha(p)\), and \(i_z(p)\) are cached cell indices; and \(R(p)\) is the composite-domain region identifier. A direct peer copy transfers the destination-contiguous buffer. GPU 1 places \(\mathcal Q_p\) in a recycled slot, if one is available, or at the end of its active range. The subsequent reorder removes the vacated source entry and incorporates the incoming particle into GPU 1's dense local array.

\subsection{Counting sort and warp aggregation}

{\color{black}
A \emph{cell key} is one integer that identifies an active three-dimensional
mesh cell.  For key $\chi$, $C_\chi$ is the number of local particles in that
cell and $O_\chi$ is the first output-array position reserved for the cell.
Counting the keys and taking an exclusive prefix sum therefore determine
where every same-cell block begins; the method is a counting sort and does
not compare particle coordinates pair by pair.
}

Consider six locally owned particles whose packed cell keys, in their current storage order, are
\[
(5,\,2,\,5,\,3,\,5,\,2).
\]
The counting pass obtains
\[
C_2=2,\qquad C_3=1,\qquad C_5=3.
\]
With keys ordered as \(2<3<5\), the exclusive offsets are
\[
O_2=0,\qquad O_3=C_2=2,\qquad
O_5=C_2+C_3=3.
\]
The scatter and gather steps therefore produce the cell-major key sequence
\[
(2,\,2,\,3,\,5,\,5,\,5).
\]
Every particle field is gathered with the same permutation, so positions, velocities, weights, cached indices, and region identifiers remain associated with the correct particle.

{\color{black}
Concretely, the two particles with key 2 occupy output positions 0 and 1,
the particle with key 3 occupies position 2, and the three particles with key
5 occupy positions 3--5.  The same six destination indices are then applied
to every structure-of-arrays component; otherwise, for example, a reordered
position could become paired with another particle's velocity.
}

The three adjacent particles with key 5 can now belong to one warp-local group \(G_5\). For a selected corner \(\bm d\), define the individual density increment
\[
\delta n_{p,\bm d}
=\frac{W_p}{V_{\nu(\bm d)}}
w_r^{(d_r)}w_\alpha^{(d_\alpha)}w_z^{(d_z)}.
\]
The conventional kernel performs one global atomic addition of \(\delta n_{p,\bm d}\) for each particle \(p\in G_5\). The aggregated kernel first computes
\[
\Delta n_{\nu(\bm d),G_5}
=\sum_{p\in G_5}\delta n_{p,\bm d}
\]
within the warp and then issues one global atomic addition for that corner. Repeating this reduction for the eight values of \(\bm d\) replaces \(3\times8=24\) global atomic additions by at most eight. The deposited value is mathematically identical, although finite-precision roundoff can differ because the summation order has changed.

{\color{black}
Here $G_5$ means only the particles with key 5 that happen to lie in the same
32-thread warp; a cell containing more than one warp is reduced once per
warp.  The binary vector $\bm d=(d_r,d_\alpha,d_z)$ selects one of the eight
cell corners, $\nu(\bm d)$ is that corner's global node, $W_p$ is particle
$p$'s macro-particle weight, and $V_{\nu(\bm d)}$ is the cylindrical control
volume of the node.  Warp aggregation changes only how equal-cell
contributions are added.  It neither drops a corner nor combines particles
from different cells.
}

\subsection{Why the candidate search uses several electron weights}

{\color{black}
This example distinguishes three quantities that are easy to confuse.
$H_e(k)$ and $H_i(k)$ are the numbers of electrons and ions in axial cell
$k$; $K$ is the number of electron steps per ion step; and $a$ is only a
temporary integer used to generate an alternative cut.  A cut $c_1=j$ puts
cells $0,\ldots,j-1$ on GPU 0 and cells $j,\ldots,N_z-1$ on GPU 1.  Every
candidate proposed with $a$ is finally judged with the physical work model
$K H_e+H_i$ and with the separate electron and ion capacities.
}

Consider \(P=2\) GPUs, \(N_z=8\) axial cells, and ion-subcycle factor \(K=4\). Let the electron and ion histograms be
\[
\begin{aligned}
H_e&=(45,40,20,20,35,5,5,5),\\
H_i&=(20,30,40,30,10,5,15,35).
\end{aligned}
\]
Equation~\eqref{eq:particle_work_histogram} then gives
\[
H_{\mathrm{work}}
=(200,190,120,110,150,25,35,55).
\]
\revgreen{For instance, the first entry is $4\times45+20=200$: the work
model counts each electron four times because it is pushed four times during
one ion-update interval.  This weighted value predicts execution time; it is
not the number of particle records stored in memory.  Electron and ion arrays
have separate fixed capacities, so a cut that balances $4H_e+H_i$ can still
place too many particles of one species on a device.  The capacity test must
therefore be applied independently to $H_e$ and $H_i$.}
Suppose that the current interior cut is \(c_1=4\). The true work loads are \(620\) and \(265\), so
\[
\eta_{\mathrm{work}}(c_1=4)
=\frac{2\max(620,265)}{620+265}
=1.401.
\]

Using the physical weight \(a=K=4\) to generate a new cut places the boundary after cell 1, i.e., \(c_1=2\). The resulting true work loads are \(390\) and \(495\), giving \(\eta_{\mathrm{work}}=1.119\). The electron loads are \(85\) and \(90\), but the ion loads are \(50\) and \(135\). Consequently,
\[
\eta_i(c_1=2)
=\frac{2\max(50,135)}{185}
=1.459.
\]
This candidate would be rejected if the prescribed worst-species limit were \(1.35\), despite its good work balance.

For the auxiliary weight \(a=1\), the proposal histogram is
\[
H_1=H_e+H_i=(65,70,60,50,45,10,20,40).
\]
Balancing this histogram produces the interior cut \(c_1=3\). Evaluating that cut with the physical weight \(K=4\) gives true work loads \(510\) and \(375\), hence \(\eta_{\mathrm{work}}=1.153\). Its electron loads are \(105\) and \(70\), and its ion loads are \(90\) and \(95\), so
\[
\eta_{\mathrm{species}}(c_1=3)
=\max\!\left(\frac{2(105)}{175},
             \frac{2(95)}{185}\right)
=1.200.
\]
The \(a=1\) candidate is therefore slightly less balanced in predicted work than the \(a=4\) candidate, but it satisfies the species limit and still improves substantially over the old work ratio. This example shows that the trial weight \(a\) generates alternatives; it never replaces \(K\) in the final work evaluation.

{\color{black}
In plain terms, the work-only cut $c_1=2$ is faster on paper but puts too many
ions on GPU 1.  Trying $a=1$ exposes the safer cut $c_1=3$; when that cut is
re-evaluated with the true factor $K=4$, it is still well balanced and no
species exceeds the chosen imbalance limit.  The multi-weight search is thus
a small discrete candidate generator, not a change to the physical time
integration.
}

The capacity and memory tests are applied after this numerical ranking. For each candidate, the algorithm checks Eq.~\eqref{eq:particle_capacity_constraint} for both species and estimates the temporary send and receive buffers produced by the cut change. Under ordinary operation, an accepted cut improves true work balance while respecting the species limit. Under capacity pressure, a strict improvement of the worst-species imbalance is the primary criterion, because avoiding particle-array exhaustion takes precedence over the smaller difference between the predicted work ratios. The selected cuts and logical-to-physical device permutation are saved in the checkpoint so that restart reconstructs the same ownership state before particle data are loaded.

\renewcommand{\bibfont}{\fontsize{7.5pt}{9pt}\selectfont}
\bibliographystyle{cas-model2-names}
\bibliography{cas-refs}

\end{document}